\documentclass[12pt]{article}

\usepackage[square,comma,sort&compress]{natbib}
\usepackage[english]{babel}
\usepackage{latexsym}
\usepackage{amsfonts}
\usepackage{amssymb} 
\usepackage{dsfont}
\usepackage{amsmath}
\usepackage{array,calc}

\numberwithin{equation}{section}

\begin{document}
\def\op#1{\mathcal{#1}}
\def\bfnull{\relax{\rm O \kern-.635em 0}}
\def\dop{{\rm d}\hskip -1pt}
\def\a{\alpha}
\def\b{\beta}
\def\g{\gamma}
\def\d{\delta}
\def\ep{\epsilon}
\def\ve{\varepsilon}
\def\t{\theta}
\def\l{\lambda}
\def\m{\mu}
\def\n{\nu}
\def\pg{\pi}
\def\r{\rho}
\def\s{\sigma}
\def\t{\tau}
\def\z{\zeta}
\def\c{\chi}
\def\p{\psi}
\def\o{\omega}
\def\G{\Gamma}
\def\D{\Delta}
\def\T{\Theta}
\def\L{\Lambda}
\def\Pg{\Pi}
\def\S{\Sigma}
\def\O{\Omega}
\def\pb{\bar{\psi}}
\def\cb{\bar{\chi}}
\def\lb{\bar{\lambda}}
\def\Pii{\mathcal{P}}
\def\Q{\mathcal{Q}}
\def\K{\mathcal{K}}
\def\A{\mathcal{A}}
\def\N{\mathcal{N}}
\def\F{\mathcal{F}}
\def\Gi{\mathcal{G}}
\def\Ci{\mathcal{C}}
\def\oL{\overline{L}}
\def\oM{\overline{M}}
\def\wk{\widetilde{K}}
\def\hb{\bar{h}}
\def\eq#1{(\ref{#1})}
\newcommand{\be}{\begin{equation}}
\newcommand{\ee}{\end{equation}}
\newcommand{\ba}{\begin{eqnarray}}
\newcommand{\ea}{\end{eqnarray}}
\newcommand{\ban}{\begin{eqnarray*}}
\newcommand{\ean}{\end{eqnarray*}}
\newcommand{\nn}{\nonumber}
\newcommand{\nin}{\noindent}
\newcommand{\fgl}{\mathfrak{gl}}
\newcommand{\fu}{\mathfrak{u}}
\newcommand{\fsl}{\mathfrak{sl}}
\newcommand{\fsp}{\mathfrak{sp}}
\newcommand{\fusp}{\mathfrak{usp}}
\newcommand{\fsu}{\mathfrak{su}}
\newcommand{\fp}{\mathfrak{p}}
\newcommand{\fso}{\mathfrak{so}}
\newcommand{\fg}{\mathfrak{g}}
\newcommand{\fr}{\mathfrak{r}}
\newcommand{\fe}{\mathfrak{e}}
\newcommand{\rE}{\mathrm{E}}
\newcommand{\rSp}{\mathrm{Sp}}
\newcommand{\rSO}{\mathrm{SO}}
\newcommand{\rSL}{\mathrm{SL}}
\newcommand{\rSU}{\mathrm{SU}}
\newcommand{\rUSp}{\mathrm{USp}}
\newcommand{\rU}{\mathrm{U}}
\newcommand{\rF}{\mathrm{F}}
\newcommand{\R}{\mathbb{R}}
\newcommand{\C}{\mathbb{C}}
\newcommand{\Z}{\mathbb{Z}}
\newcommand{\Hb}{\mathbb{H}}
\def\oL{\overline{L}}
\def\mW{\mathcal{W}}
\def\bul{\bullet}
\newcommand{\rf}[1]{(\ref{#1})}
\newcommand{\cm}[1]{{\textbf{#1}}}
\newcommand{\pt}{\partial}
\newcommand{\pts}{\slash{\partial}}
\newcommand{\sh}{\slash{H}}
\newcommand{\sph}{\slash{\Phi}}
\newcommand{\sps}{\slash{\Psi}}

\newcommand{\Cset}{\mathbb{C}}
\newcommand{\CY}{{\mathrm{CY}}}
\newcommand{\e}{{\mathrm e}}
\newcommand{\eg}{{\it e.g.}~}
\newcommand{\ie}{{\it i.e.}\ }
\newcommand{\Nset}{\mathbb{N}}
\newcommand{\Pset}{\mathbb{P}}
\newcommand{\Pin}{\mathrm{Pin}}
\newcommand{\Qset}{\mathbb{Q}}
\newcommand{\Rset}{\mathbb{R}}
\newcommand{\SL}{\mathrm{SL}}
\newcommand{\sign}{\mathrm{sign}}
\newcommand{\SO}{\mathrm{SO}}
\newcommand{\mO}{\mathrm{O}}
\newcommand{\Spin}{\mathrm{Spin}}
\newcommand{\SU}{\mathrm{SU}}
\newcommand{\tD}{\tilde{D}}
\newcommand{\tg}{\tilde{\gamma}}
\newcommand{\tr}{\mathrm{tr}}
\newcommand{\Tset}{\mathbb{T}}
\newcommand{\U}{\mathrm{U}}
\newcommand{\unit}{\mathbf{1}}
\newcommand{\Vol}{\mathrm{Vol}}
\newcommand{\Zset}{\mathbb{Z}}

\def\a{\alpha}
\def\b{\beta}
\def\g{\gamma}
\def\c{\chi}
\def\d{\delta}
\def\e{\epsilon}
\def\ep{\varepsilon}
\def\f{\phi}
\def\vf{\varphi}
\def\z{\psi}
\def\zb{\overline{\psi}}
\def\zt{\widetilde{\psi}}
\def\k{\kappa}
\def\l{\lambda}
\def\m{\mu}
\def\n{\nu}
\def\o{\omega}
\def\q{\theta}
\def\th{\theta}
\def\tc{\hat{\theta}}
\def\r{\rho}
\def\s{\sigma}
\def\st{\widetilde{\sigma}}
\def\sut{\utw{\sigma}}
\def\t{\tau}
\def\u{\upsilon}
\def\x{\xi}
\def\z{\zeta}
\def\w{\wedge}
\def\D{\Delta}
\def\F{\Phi}
\def\G{\Gamma}
\def\J{\Phi}
\def\L{\Lambda}
\def\O{\Omega}
\def\P{\Pi}
\def\Q{\Theta}
\def\X{\Xi}

\def\cN{{\cal N}}
\def\cP{{\cal P}}

\newcommand{\bea}{\begin{eqnarray}}
\newcommand{\eea}{\end{eqnarray}}
\newcommand{\ft}[2]{{\textstyle\frac{#1}{#2}}}
\def\fft#1#2{{\frac{#1}{#2}}}
\def\ul#1{\underline{#1}}
\newcommand{\half}{\frac{1}{2}}

\def\slashchar#1{\setbox0=\hbox{$#1$}           
\dimen0=\wd0                                 
\setbox1=\hbox{/} \dimen1=\wd1               
\ifdim\dimen0>\dimen1                        
\rlap{\hbox to \dimen0{\hfil/\hfil}}      
#1                                        
\else                                        
\rlap{\hbox to \dimen1{\hfil$#1$\hfil}}   
/                                         
\fi}

\def\slash#1{\rlap {\begin{picture}(10,10) \put(0,0){\line(1,1){10}} \end{picture}} #1}

\def\slas#1{\rlap{\begin{picture}(10,10)(-5,0)
\put(0,0){\line(2,1){15}}
\end{picture}} #1 }

\def\slashh#1{\rlap{\begin{picture}(10,10)
\put(0,0){\line(5,1){40}}
\end{picture}}#1}

\newcommand{\sla}{\slash\!\!\!\!}
\def\sei{e^{i J}\!\!\!\!\! 
\begin{picture}(10,10)
\put(0,0){\line(1,2){5}}
\end{picture}
}


\begin{titlepage}
\begin{center}

\rightline{\small IFT-UAM/CSIC-09-24}
\rightline{\small ZMP-HH/09-11}
\vskip 1cm

{\Large \bf Domain wall flow equations and $SU(3)\times SU(3)$ structure compactifications}
\vskip 1.2cm
	
{\bf 
Paul Smyth$^1$ and Silvia Vaul\`a$^{2}$ }
\vskip 0.2cm
$^1$\textit{ II. Institut f\"ur Theoretische Physik der
Universit\"at Hamburg\\
Luruper Chaussee 149, 22761 Hamburg, Germany} \\
{\small\upshape\tt paul.smyth@desy.de }\\[3mm]
$^2$ {\it Instituto de F\'{\i}sica Te\'orica UAM/CSIC\\
Facultad de Ciencias C-XVI,  C.U.~Cantoblanco,  E-28049-Madrid, Spain} \\
{\small\upshape\tt silvia.vaula@uam.es} 
\vskip 0.4cm

\end{center}

\vskip 1cm

\begin{center}
\textit{Dedicated to the memory of Raffaele Punzi}
\end{center}
 
\vskip 1cm

\begin{center} {\bf Abstract }\end{center}
\vskip 0.4cm

\noindent We study supersymmetric domain wall solutions in four dimensions arising from the compactification of type II supergravity on an $SU(3) \times SU(3)$ structure manifold. Using a pure spinor approach, we show that the supersymmetry variations can be reinterpreted as a generalisation of the Hitchin flow equations and describe the embedding of an $SU(3) \times SU(3)$ structure manifold into a $G_2 \times G_2$ structure manifold. We find a precise agreement between the four- and ten-dimensional supergravity results. The flow equations derived here should have applications in constructing the gravity duals of Chern-Simons-matter conformal field theories. 

\vfill

\end{titlepage}

\tableofcontents

\newpage

\section{Introduction}\label{intro}

The task of finding compactifications of ten-dimensional supergravity that lead to a four-dimensional effective theory with $\cN=1$ supersymmetry has received much attention. One promising approach is compactification with flux, where the internal manifold can be deformed away from being Ricci-flat (see, for instance, \cite{Grana:2005jc} for a review). Fluxes can induce torsion, meaning that the internal six-manifold will no longer have $SU(3)$ holonomy, for example, but rather $SU(2)$ or $SU(3)$ structure. The constraints on the geometry of the internal manifold are then most conveniently rephrased in terms of generalised complex geometry using $SU(3)\times SU(3)$ structures defined on the formal sum of the tangent and cotangent bundles \cite{Hitchin:2004ut,Gualtieri:2003dx,gmpt1,Witt,gmpt2,gmpt3,Grana:2005ny,Grana:2006hr}. Finding explicit examples of such manifolds that lead to four-dimensional Minkowski vacua has proven difficult in practice \cite{gmpt3}. This led to the search for alternative $AdS_4$ vacua (see \cite{Tomasiello:2007eq,Kounnas:2007dd,Koerber:2008rx} and references therein) which may also provide a useful starting point for realistic models via the KKLT proposal \cite{Kachru:2003aw}. In this paper we shall focus on another class of four-dimensional vacuum configurations - domain walls - which are readily found in gauged supergravity \cite{Mayer:2004sd,Louis:2006wq,Behrndt:2001mx}. The near-horizon limit of a domain wall in four dimensions can produce an $AdS_4$ spacetime that can be interpreted as arising from an $SU(3)\times SU(3)$ structure reduction of type II supergravity \cite{Kounnas:2007dd,Koerber:2008rx}. Much of the work to date has focused on the construction of examples of $AdS_4$ vacua that can be reinterpreted as stacks of orthogonally intersecting branes and Kaluza-Klein monopoles in ten dimensions. Domain wall probes in $SU(3)\times SU(3)$ structure backgrounds have also been studied for their interesting supersymmetry-breaking properties \cite{Lust:2008zd}. The aim of this article is to provide a general characterisation of domain wall \textit{vacua} arising from $SU(3)\times SU(3)$ structure compactifications. 

The equations of motion for supersymmetric domain wall configurations are known to reduce to a set of first-order flow equations for scalar fields \cite{Cvetic:1996vr}. This is exemplified in four-dimensional, matter-coupled, $\cN=2$ supergravity arising from type IIA supergravity compactified on a half-flat six-manifold. For a domain wall configuration one finds a set of first-order flow equations for the hyper and vector multiplet scalar fields. This system of equations can be shown to be equivalent to the Hitchin flow equations \cite{Mayer:2004sd}, which describe the embedding of the half-flat SU(3) structure six-manifold into a $G_2$-holonomy seven-manifold with boundary \cite{HitchinHF}. The mirror symmetric  description of this configuration is given by type IIB supergravity compactified on a Calabi-Yau six-manifold with electric Neveu-Schwarz (NS) fluxes. This was further extended to type IIB supergravity with electric and magnetic NS fluxes in \cite{Louis:2006wq}, which is mirror dual to type IIA supergravity compactified on a manifold with $SU(3)\times SU(3)$ structure. The flow equations derived in four dimensions were shown to be equivalent to the generalised Hitchin flow equations \cite{Witt,Jeschek}, which describe the embedding of an $SU(3)\times SU(3)$ structure manifold into $G_2\times G_2$, or generalised $G_2$ \cite{Witt2}, structure manifold. We shall further extend this analysis to a more general set of four-dimensional charges comprising torsion, Ramond-Ramond (RR), NS  and non-geometric fluxes, and correspondingly derive a general set of flow equations in four and ten dimensions. Our results can be interpreted as describing the embedding of an $SU(3)\times SU(3)$ structure manifold into an `almost' $G_2\times G_2$ structure manifold, with the RR fields providing the obstruction to integrability of the generalised almost-$G_2$ structure.

The dual approach of analysing the domain wall flow equations in both four and ten dimensions has the advantage of circumventing some well-known problems. One of the outstanding technical issues in flux compactifications is to find an appropriate definition of the spectrum of light modes in the absence of harmonic forms \cite{Grana:2005ny,Grana:2006hr,KashaniPoor:2007tr}. While progress has been made for coset- and nil-manifolds \cite{Caviezel:2008ik}, the comparison of the truncated four-dimensional theory and its ten-dimensional counterpart is complicated and unclear in general. On the other hand, from the four-dimensional gauged supergravity point of view, the models corresponding to flux compactifications  \cite{Dall'Agata:2003yr,Sommovigo:2004vj}, $SU(3)$ structure compactifications with electric RR and NS fluxes  \cite{D'Auria:2004tr} and $SU(3)\times SU(3)$ structure compactifications with electric and magnetic RR and NS fluxes  \cite{D'Auria:2007ay} are well defined theories. In order to have a more complete understanding of generalised compactifications and the various low-energy vacua, it is worthwhile to study the problem in both four and ten dimensions. 

We shall initially follow the completely four-dimensional approach and look for half-supersymmetric, BPS, domain wall solutions of general gauged $\cN=2$ supergravity. In section \ref{s2} we derive the domain wall flow equations, following the analysis of \cite{Louis:2006wq,Behrndt:2001mx,LopesCardoso:2001rt}, and further discuss their modification in the presence of orientifold projections. In section \ref{s3} we shall derive the flow equations from the ten-dimensional perspective. Our goal will be to find a set of equations describing domain wall vacua of SU(3)$\times$SU(3) structure compactifications by manipulating the ten-dimensional type II supersymmetry transformations, inspired by the black hole discussion in \cite{hmt}. For simplicity, we choose to focus on configurations preserving {\it at least} $1/16$-supersymmetry i.e. at least two unbroken supercharges. This allows us to follow the calculation of the $\cN=1$, Minkowski and $AdS_4$ vacuum conditions described in \cite{gmpt3}, appropriately modified for domain wall spacetimes, and write our results in terms of pure spinors $\Phi_+$ and $\Phi_-$. The resulting expressions will provide an extension of the results of \cite{Witt} to compactifications with non-trivial RR fields, as well as more general domain wall profiles (see also \cite{hlmt} for related work). In section \ref{s4} we compare our two results and show that there is a precise agreement between the four- and ten-dimensional derivations. For clarity, we provide some simple examples with vanishing RR fluxes and we also discuss the extension of our results to a non-geometric setting. The generalised Hitchin flow equations that we find are consistent with the proposal of generalised mirror symmetry \cite{gmpt2}, interchanging the IIA/IIB fluxes and the two pure spinors $\Phi_+ \leftrightarrow \Phi_-$. Furthermore, our results provide a useful on-shell check of the truncation and reduction proposal of \cite{Grana:2006hr}.  We present our conclusions and discuss possible applications of our results in section \ref{disc}.

The reader should be aware that in order to more easily make contact with the literature, we have chosen to use different metric conventions in four and ten dimensions, and that an explanation of the dictionary can be found in section \ref{s4}. Due to this technicality, we have provided a pedagogical review of our conventions in appendix \ref{conv}. Appendices \ref{psd} and \ref{fermtrans} contain a review of pure spinors and the necessary features of general four-dimensional, matter-coupled, $\cN=2$ supergravity. Appendix \ref{4dstr} specialises to  theories arising from type II supergravity compactified on $SU(3)\times SU(3)$ structure manifolds with electric and magnetic RR and NS fluxes. In appendix \ref{DWmanipul} we provide details of the derivation of the Hitchin flow equations from the four-dimensional $\cN=2$ supersymmetry transformations for $SU(3)\times SU(3)$ structure compactifications.


\section{Domain wall vacua in 4D supergravity}\label{s2}

\subsection{4D $\cN=2$ supergravity}

Let us consider an off-shell compactification of type IIA supergravity on an $SU(3)\times SU(3)$ structure manifold \cite{Grana:2006hr,Cassani:2009ck} $\hat{Y}$, with an $SU(3)$ structure manifold $Y$ \cite{Grana:2005ny} as a particular case. The resulting theory is a four-dimensional, $\cN=2$ gauged supergravity coupled to $n_V=h^{(1,1)}$ vector multiplets and $n_H=(h^{(2,1)}+1)$ scalar/tensor hypermultiplets \cite{Dall'Agata:2003yr,Sommovigo:2004vj}. The most important point for us is the relation between the fields of type IIA supergravity and those of the four-dimensional theory. When compactifying on a Calabi-Yau manifold, one is accustomed to making a harmonic expansion of the various fields and truncating to the set of modes which are massless in four dimensions. Motivated by this, \cite{Grana:2005ny,Grana:2006hr} proposed that in $SU(3)$ and $SU(3)\times SU(3)$ compactifications, where the distinction between heavy and light modes is unclear due to the lack of a harmonic expansion, one should proceed by truncating the space of forms to a finite-dimensional subspace. The guiding physical principle was that one should aim to be left with only the gravitational multiplet along with vector, tensor and hypermultiplets. In particular, all possible spin-$3/2$ multiplets should be projected out. Furthermore, the truncation should not break supersymmetry and, therefore, the special K\"ahler metrics on the moduli spaces of the pure spinors $\Phi_\pm$, which are in one-to-one correspondence with metric deformations, should descend to the truncated subspaces. One then defines a set of basis forms on these subspaces, in terms of which the ten-dimensional fields and the pure spinors can be expanded. We shall now review our conventions for the truncated pure spinors, which we denote $\Phi^0_\pm$ and $\hat\Phi^0_\pm$ in the $SU(3)$ and $SU(3)\times SU(3)$ structure cases, respectively. We refer the reader to appendix \ref{4dstr} for further details. 

When $Y$ has $SU(3)$ structure the truncated pure spinors $\Phi^0_\pm$ are defined as 
\ba
&&\Phi^0_+=X^{\bf\L}\o_{\bf\L}-F_{\bf\L}\o^{\bf\L}\label{Phi0+}~,\\
&&\Phi^0_-=Z^A_\eta\a_A-G_{\eta\,A}\b^A\label{Phi0-}~,
\ea
where ${\bf\L}=(0,\,i)$, $i=1,\dots h^{(1,1)}$ and $A=(0,\,a)$, $a=1,\dots h^{(2,1)}$. The non-harmonic basis of forms \eq{asu3basis} on $Y$ is
\be 
(\o_{\bf\L}\,,\,\o^{\bf\L}),\quad\quad (\a_A\,,\,\b^A)~.\label{su3basis}
\ee 
The rescaled sections introduced in \eq{Phi0-} are defined as
\be
(Z^A_\eta\,,G_{\eta A})\equiv \eta (Z^A\,,G_A)\label{Zresc}~,
\ee
where $\eta$ is a normalisation factor (see the discussion around \rf{eta} for more details). $X^{\bf\L}$ and $F_{\bf\L}=\frac{\partial F}{\partial X^{\bf\L}}$ are the homogeneous complex coordinates and the derivative of the holomorphic prepotential $F$ for the K\"ahler moduli, respectively. $Z^A$ and $G_A=\frac{\partial G}{\partial Z^A}$ are the homogeneous complex coordinates and the derivative of the holomorphic prepotential $G$ for the complex structure moduli, respectively.

The failure of the basis forms \rf{su3basis} to be closed can be expressed as
\ba 
d_H\a_A=e_{A{\bf\L}}\o^{\bf\L},\quad\quad  d_H\b^A=e^A_{\bf\L} \o^{\bf\L}~,\nn\\
d_H\o_{\bf\L}=e_{\bf\L}^A\a_A-e_{A{\bf\L}}\b^{A},\quad\quad d_H\o^{\bf\L}=0\label{decohSU3TH}~,
\ea
where it is convenient to define a twisted derivative operator $ d_H\equiv d-H\wedge$. The NS flux gives rise to
\be
\label{NSflux}H=e_0^A{\a}_A-e_{0A}\b^A~,
\ee
which we shall call $H$ deformations. We refer to the remaining flux parameters $(e^A_i,\,e_{Ai})$ as $T$(orsion) deformations, as they all have a geometric origin. Finally, we note that $d^2=d_H^2=0$ implies
\be 
e_{A{\bf\L}}e^A_{~\bf\S}-e_{A{\bf\S}}e^A_{~\bf\L}=0\label{notadSU3}~.
\ee

The RR fluxes are introduced using the basis of even forms \eq{su3basis}, according to
\be 
F^{flux}=e_{\bf\L}\o^{\bf\L}-m^{\bf\L}\o_{\bf\L} \label{RRf3}.
\ee

The $SU(3)\times SU(3)$ structure case is somewhat more complicated. We shall review the main points here and refer the reader to appendix D for further explanations. When $\hat Y$ has $SU(3)\times SU(3)$ structure we define a basis of polyforms \eq{aPFbasis} as
\be 
(\hat\o_{\bf\L},\,\hat\o^{\bf\L})\quad {\bf\L}=0,1,\dots h^{(1,1)}\quad , \quad  (\hat\a_A\,,\,\hat\b^A)\quad A=0,1,\dots h^{(2,1)}~,\label{PFbasis}
\ee
i.e. the basis forms are no longer of fixed degree. The truncated pure spinors $\hat\Phi^0_\pm$ are then defined as
\ba
&&\hat\Phi_+^0=X^{\bf\L}\hat\o_{\bf\L}-F_{\bf\L}\,\hat\o^{\bf\L}\label{p1}~,\\
&&\hat\Phi_-^0=Z^A_\eta\hat\a_A-G_{\eta A}\,\hat\b^A\ .\label{p2}
\ea
To discuss the non-closure of the basis forms \rf{PFbasis} it is convenient to introduce a {\it generalised differential} (see e.g. \cite{Grana:2006hr} and references therein)  
\be 
\mathcal{D}\equiv d\ -H\wedge\ \,-Q\cdot\ \,-R\llcorner\label{defD}~,
\ee
where $Q\cdot\ $ and $R\llcorner\ $ act on a generic $k$-form $C$ as
\be 
(Q\cdot C)_{m_1\dots m_{k-1}}=Q^{ab}_{\ \ [m_1}C_{ab m_2\dots m_{k-1}]},\quad\quad (R\llcorner C)_{m_1\dots m_{k-3}}=R^{abc}C_{abcm_1\dots m_{k-3}}~.
\ee
The action of $\mathcal{D}$ on the basis forms and the subsequent constraints on the fluxes is given in \rf{decohSU33}-\rf{2notadSU33}. 

The RR fluxes are introduced using the same basis \eq{PFbasis}
\be 
\hat{F}^{flux}=e_{\bf\L}\hat{\o}^{\bf\L}-m^{\bf\L}\hat{\o}_{\bf\L} \label{RRf33}\ .
\ee
When considering the RR fields, it is useful to introduce a third polyform $\hat\S$  (that reduces to a three-form $\S$  in ${\rm SU}(3)$ structure case)
\be
\hat\S=\zeta^A\hat\alpha_A-\tilde\zeta_{ A}\hat\beta^A~,\label{p3}
\ee
such that the total RR contribution is given by
\be
\hat{F}=\mathcal{D}\hat{\S}+\hat{F}^{flux}~. \label{totalRRf}
\ee
In the ${\rm SU}(3)$ structure case this becomes
\be 
F=d_H\S+F^{flux}\label{totalRRff}\ .
\ee
The fields $(\z^A,\tilde\z_A)$ are the RR scalars or dual tensors in the hypermultiplet sector in four dimensions. One may notice that the definition \eq{p3} is sensitive to the fact that some RR fields may appear as tensors in four dimensions in the  $SU(3)\times SU(3)$ structure case. This is not a problem, as in all expressions of interest $\hat\S$ only appears through its generalised derivative $\mathcal{D}\hat\S$ \eq{totalRRf}, which contains just the gauge invariant combinations that are not dualised into tensors \cite{D'Auria:2007ay}. Finally, we note that it is often convenient to introduce the rescaled RR fields
\be 
(\z^A_{\tilde\eta},\,\tilde\z_{A\tilde\eta} )\equiv \tilde\eta (\z^A,\,\tilde\z_A)\label{Srescalintro}~,
\ee
and the rescaled total RR contribution 
\be
\hat{F}_{\tilde\eta}\equiv \tilde\eta\hat{F}\label{Frescintro}~.
\ee
where $\tilde\eta$ is a normalization factor (see the discussion around \eq{ap3}).


Let us now return to the four-dimensional, $\cN=2$ gauged supergravity. The gravitational multiplet is given by
\be(g_{\m\n},\,\p_{\hat{A}\m},\,\p^{\hat{A}}_\m,\,A^0_\m)~,\ee
where $g_{\m\n}$ is
the metric, $\p_{\hat{A}\m},\, \hat{A}=1,2$ are the two chiral gravitini and $A^0_\m$
is the graviphoton. The vector multiplets are given by
\be(A^i_\m,\,\l^{i\hat{A}},\,\l^i_{\hat{A}},\,t^i)\ ,\qquad i = 1,\ldots, n_V\ ,\ee 
where 
$A^i_\m$ are  the gauge bosons, $\l^{i\hat{A}}$ are the doublets of chiral gaugini and $t^i$ are the complex scalar fields appearing in the expansion of truncated pure spinors\footnote{We refer the reader to \cite{Grana:2005ny,Grana:2006hr} for a thorough discussion of the reduction and truncation of the type II theories to four-dimensional {\cN}=2 supergravity.} $\Phi^0_+$ and $\hat\Phi^0_+$ \eq{Phi0+}, \eq{p1}. The $n_H$ hypermultiplets contain two chiral hyperini, which we collectively denote as $\zeta_{\hat\alpha}$, and four bosons, which in principle can be scalar $q^u$ or tensor $B_{I\mu\nu}$ fields,
\ba (\zeta_{\hat\alpha},\,\zeta^{\hat\alpha},\,q^u,\,B_{I\m\n})\ , && \quad\hat\alpha =
1,\ldots,2n_H+2\ , \\
&&\quad  u = 1,\ldots, 4n_H+4-n_T,\quad I=1,\dots n_T\ .\nn \ea
If the tensors are massless, e.g. as happens for the universal hypermultiplet tensors in Calabi-Yau compactifications, they can be dualised into scalar fields; otherwise, they have to be kept as massive tensors. The scalar fields $q^u$ contain the dilaton $\varphi$,  the complex scalars $z^a$ appearing in the expansion of the truncated pure spinors $\Phi^0_-$ and $\hat\Phi^0_-$ \eq{Phi0-}, \eq{p2}, and the scalars $(\z^A,\,\tilde\z_A)$ appearing in reduction of the RR sector.  In the $SU(3)$ structure case \cite{D'Auria:2004tr},  $(\z^A,\,\tilde\z_A)$ appear as scalars in the four-dimensional theory, while in the $SU(3)\times SU(3)$ structure case \cite{D'Auria:2007ay} some combinations are dualised into  tensors. The four-dimensional components of the NSNS two-form $B_{\m\n}$ naturally appear as a tensor. Nevertheless, in the $SU(3)$ structure case it is massless and for convenience it is dualised into a scalar. 

For the domain wall configurations that we are interested in, we only consider the situation where both the vector and tensor field strengths are vanishing.  With this assumption the supersymmetry transformation laws for the fermions \eq{ferm} simplify and the supersymmetry conditions are
\ba
\d\p_{\m \hat{A}}&=&D_\m\ve_{\hat{A}}+iS_{\hat{A}\hat{B}}\g_\m\ve^{\hat{B}}=0\label{psi}~, \\
\d\l^{i\hat{A}}&=&i\partial_\m t^i\g^\m\ve^{\hat{A}}+W^{i\hat{A}\hat{B}}\ve_{\hat{B}}=0\label{la}~,\\
\d\z_{\hat\a}&=&iP_{u \hat{A}\hat\a}\partial_\m q^u\g^\m\ve^{\hat{A}}+N_{\hat\a}^{\hat{A}}\ve_{\hat{A}}=0\label{zi}~.
\ea
Here  $P_{u {\hat{A}}\hat\a}$ parameterise the scalars in the hypermultiplets. Note that in the absence of (hyper) tensor multiplets,  $P_{u {\hat{A}}\hat\a}$ coincide with the vielbein of the quaternionic manifold. In the presence of  (hyper) tensor multiples, they parameterise the scalars which have not been dualised into tensors \cite{Dall'Agata:2003yr}. The fermion shifts $S_{{\hat{A}}{\hat{B}}}$, $W^{i{\hat{A}}{\hat{B}}}$ and $N_{\hat\a}^{\hat{A}}$ encode the gauging.  For the cases we are interested in, the gauging will always be Abelian \cite{Sommovigo:2004vj,D'Auria:2004tr,D'Auria:2007ay}. As a consequence, the electric and magnetic  triholomorphic momentum map takes the simple form
\be\vec P_{\bf\L}=\vec\o_I k^I_{\bf\L},\quad\quad \vec Q^{\bf\L}=\vec\o_I k^{I\bf\L}~,\ee
where $k^I_{\bf\L}$ are the Killing vectors associated to the gauge group and $k^{I\bf\L}$ are their magnetic duals. Using the homogeneous special coordinates $X^{\bf\L}$ and the derivative of the holomorphic prepotential $F_{\bf\L}$ for the K\"ahler moduli space, defined in appendix \ref{4dstr},  we can define an $SU(2)$ triplet of  superpotentials 
\be
\vec W=e^{\frac{K_+}2}\left(X^{\bf\L} \vec{P}_{\bf\L}-F_{\bf\L} \vec{Q}^{\bf\L}\right)~,\label{SU2W}
\ee
such that the fermion shifts are related to the superpotentials $\vec W$ as follows
\ba
\label{S}S_{{\hat{A}}{\hat{B}}}&=&\frac i2 \vec\s_{{\hat{A}}{\hat{B}}}\cdot\vec W~,\\
\label{Wi}W^{i{\hat{A}}{\hat{B}}}&=&i\vec\s^{{\hat{A}}{\hat{B}}}g^{i\bar{\jmath}}\nabla_{\bar\jmath}\vec W~,\\
\label{N}P^{v{\hat\a}}_{({\hat{A}}}N_{{\hat{B}})\hat\a} &=&i\vec\s_{{\hat{A}}{\hat{B}}}h^{vu}\nabla_u\vec W~,
\ea
where $g_{i\bar{\jmath}}$ and $h_{uv}$ are the metrics of the scalar $\s$-model in the vector and hypermultiplet sectors, respectively, and $K_+$ is the K\"{a}hler potential defined in  \eq{JJJ} and \rf{genkp}. It is convenient to decompose the $SU(2)$ vector $\vec W$ into its norm and a unit norm vector $\vec n$ \cite{LopesCardoso:2001rt}
\be \vec W=W\,\vec n,\quad\quad \vec n\cdot\vec n=1\label{normphase}~.\ee
Multiplying \eq{normphase} by $\vec n$ we obtain the expression for the ``superpotential'' 
\be W=\vec n\cdot\vec W~.\ee

If we consider type IIA supergravity compactified on an $SU(3)$ structure manifold with electric RR fluxes\footnote{The constants $c_{\bf\L}$ in \cite{D'Auria:2004tr} are related to the electric RR fluxes by $c_{\bf\L}=-2e_{\bf\L}$.}  $e_{\bf\L}$, the corresponding gauging of the $\cN=2$ supergravity  \cite{D'Auria:2004tr} is purely electric, i.e. it corresponds to the choice $\vec{Q}^{\bf\L}=0$. The components of the $SU(2)$ superpotential $\vec W$ are given by
\ba
&&W^1= 2\, e^\varphi e^{\frac{K_++K_-}2} X^{\bf\L} {\rm Re}(G_A e^A_{~{\bf\L}}-Z^A e_{A{\bf\L}})\nn~,\\ 
&&W^2=2\, e^\varphi e^{\frac{K_++K_-}2}X^{\bf\L} {\rm Im}(G_A e^A_{~{\bf\L}}-Z^A e_{A{\bf\L}})\nn~, \label{WSU3}\\ 
&&W^3=e^{2\varphi} e^{\frac{K_+}2}X^{\bf\L} (\tilde\z_Ae^A_{~{\bf\L}}-\z^A e_{A{\bf\L}}-2e_{\bf\L})~,
\ea
where $\varphi$ is the four-dimensional dilaton, related to the ten-dimensional dilaton by
\be 
e^{2\phi}=\frac18 e^{2\varphi-K_+}\label{10Adil}~.
\ee
$K_-$ is the K\"{a}hler potential defined in \eq{OO}and \rf{genkp}. Using equations \eq{dPhi+}, \eq{dPhi-} and \eq{Ftorsflux} we can rewrite the components of the superpotential $\vec W$ in terms of the pure spinors \rf{Phi0+},  \rf{Phi0-} and the RR fields  \rf{totalRRff}  as follows
\ba
&&W^1=-2\, e^{\varphi+K_+}\langle d_H {\rm Re}\Phi^0_-,\,\Phi^0_+\rangle~,\\
&&W^2=-2\, e^{\varphi+K_+}\langle d_H {\rm Im}\Phi^0_-,\,\Phi^0_+\rangle~,\\ 
&&W^3=-2e^{\varphi+K_+}\langle F_{\tilde\eta},\,\Phi^0_+\rangle~.
\ea
$\langle \cdot,\cdot \rangle$ is the Mukai pairing \rf{Mukai} and we have dropped the $\int_Y$ from each bracket to keep our formula compact.

If we consider the more general $SU(3)\times SU(3)$ structure case with electric and magnetic RR fluxes  $\left( e_{\bf\L},\, m^{\bf\L}\right)$, the theory will contain massive tensor multiplets and the components of the superpotential $\vec W$ are given by \cite{D'Auria:2007ay}
\ba
W^1&\!\!=\!\!& 2\, e^\varphi e^{\frac{K_++K_-}2}\left[X^\L {\rm Re}(G_A e^A_{~\L}-Z^A e_{A\L})-F_\L {\rm Re}(G_A m^{A\L}-Z^A m_{A}^{~\L})\right]\nn~,\\ 
W^2&\!\!=\!\!&2\, e^\varphi e^{\frac{K_++K_-}2}\left[ X^\L {\rm Im}(G_A e^A_{~\L}-Z^A e_{A\L})-F_\L {\rm Im}(G_A m^{A\L}-Z^A m_{A}^{~\L})\right]\nn~,\\ 
W^3&\!\!=\!\!&e^{2\varphi} e^{\frac{K_+}2}\left[X^\L (\tilde\z_Ae^A_{~\L}-\z^A e_{A\L}-2e_\L)-F_\L (\tilde\z_Am^{A\L}-\z^A m_{A}^{~\L}-2m^\L)\right]\nn~.\\
&&\label{WSU33}
\ea
Using equations \eq{dPhi+}, \eq{dPhi-} and \eq{Ftorsflux} we can rewrite these expressions in terms of the pure spinors \eq{p1}, \eq{p2} and the RR fields  \rf{totalRRf}
\ba
&&\label{33w1}W^1=-2\, e^{\varphi+K_+}\langle \mathcal{D}  {\rm Re}\hat\Phi^0_-,\,\hat\Phi^0_+\rangle~,\\
&&\label{33w2}W^2=-2\, e^{\varphi+K_+}\langle \mathcal{D}  {\rm Im}\hat\Phi^0_-,\,\hat\Phi^0_+\rangle~,\\ 
&&\label{33w3}W^3=-2e^{\varphi+K_+}\langle \hat F_{\tilde\eta},\,\hat\Phi^0_+\rangle~.
\ea


\subsection{Domain wall solutions}

We are interested in domain wall spacetimes in four dimensions described by a metric of the form\footnote{In four dimensions we use the mostly-minus metric signature in order to  be consistent with references \cite{D'Auria:2004tr,D'Auria:2007ay}.}
\begin{equation}\label{4dmetric}
ds_4^2  = e^{2U(r)}\eta_{\a\b}dx^\a dx^\b -e^{-2pU(r)}dr^2 ~,
\end{equation}
where $p$ is a constant and, for simplicity, we have chosen a flat worldvolume metric $\eta_{\a\b}$, and $\a,\b=0,1,2$. For the coordinate transverse to the domain wall, we will use `$r$' and `3' to denote a curved and flat index, respectively. We shall make the usual, physically motivated assumption that scalar fields will only depend on the direction transverse to the domain wall. 

We can now proceed to analyse the supersymmetry variations as in \cite{Louis:2006wq,LopesCardoso:2001rt}. From  $\delta\p_{\a {\hat{A}}}=0$ we obtain
\be U^\prime e^{pU}\g_3\ve_{\hat{A}}=-2i\, n_{{\hat{A}}{\hat{B}}}W\ve^{\hat{B}}\label{psialpha}~,\ee
where we have used that the domain wall is flat \eq{metric} and we have defined $n_{{\hat{A}}{\hat{B}}}=\frac i2 \vec n\cdot \vec\s_{{\hat{A}}{\hat{B}}}$. The consistency of \eq{psialpha} implies
\be e^{2pU}\left(U^\prime(r)\right)^2=|W|^2\quad\Rightarrow\quad e^{pU}U^\prime(r)=\pm|W|\label{Uprime}~.\ee
Inserting \eq{Uprime} back into $\d\p_{\a {\hat{A}}}=0$ we obtain the BPS projector 
\be \ve_{\hat{A}}=\pm\hb\,\vec n\cdot\vec\s_{{\hat{A}}{\hat{B}}}\,\g_3\ve^{\hat{B}}\label{project}~,\ee
where $\hb$ is the ${\rm U}(1)_V$ phase of $W$
\be W=\hb|W|\label{Wphase}~.\ee
The subscript ``${}_V$" refers to the line bundle of the special K\"ahler geometry of the vector multiplet scalars. From $\d\p_{r{\hat{A}}}=0$ we obtain
\ba
 \hb \partial_rh&=&\pm i\, e^{-pU}{\rm Im}(hW)\label{Dh}~,\\
  \partial_r n_{{\hat{A}}{\hat{B}}}&=&0\label{Dn}~.
 \ea
For a constant curvature metric we have that \cite{LopesCardoso:2001rt}
\be D_\a (h^{\frac12}\ve_{\hat{A}})=\frac i\ell \g_\a \g_3\,h^{\frac12}\ve_{\hat{A}}~,\ee 
which can be used in  $\d\p_{r{\hat{A}}}=0$ in order to obtain
\ba
\frac1\ell&=&\pm\frac 12 e^U {\rm Im}(hW)\label{1ell}~,\\
U^\prime&=&\pm e^{-pU}{\rm Re}(hW)\label{Uprime2}~,
\ea
where $\ell^{-2}$ is proportional to the curvature. Comparing \eq{Uprime} with  \eq{Uprime2}, or alternatively taking the limit $\ell\rightarrow \infty$ in \eq{1ell}, we obtain 
\be{\rm Im}(hW)=0\label{reW}~,\ee  
which implies from \eq{Dh} that $\partial_r h=0$. From \eq{psi} we further obtain the expression for the  Killing spinor
\be \ve_{\hat{A}} (r)=e^{\frac12 U}\ve_{\hat{A}}^0 \label{4drsoln}~,\ee
where $\ve_{\hat{A}}^0$ is a constant spinor obeying the projection condition \eq{project}. Finally, using the projector \eq{project} in equations \eq{la} and \eq{zi} we obtain the following set of flow equations
\ba
&&\partial_r t^i=\mp e^{-pU}g^{i\overline\jmath}\,\hb\,\nabla_{\overline\jmath}\bar{W}\label{attz0}~,\\
&&\partial_rq^u=\mp e^{-pU} g^{uv}\,\hb\, \partial_v\bar{W}\label{attq0}~,\\
&& U^\prime=\pm e^{-pU}hW~.\label{attu}
\ea
The detailed derivation of the generalised Hitchin flow equations from \eq{attz0}-\eq{attu} is presented in appendix \ref{DWmanipul}. Here we shall only summarise the relevant results.

By manipulating \eq{attz0} we can derive the expression for the flow equation of ${\rm Im}(h\hat\Phi^0_+)$. Under the assumption that $\vec n$ does not depend on the vector multiplet scalars, we can put \eq{attz0} into the desired form  
\be
\partial_r\begin{pmatrix} {\rm Im}(h\,e^{U+\frac{K_+}2}X^{\bf\L})\cr {\rm Im }
(h\,e^{U+\frac{K_+}2}F_{\bf\L}) \end{pmatrix} =-\frac12e^{(1-p)U}\begin{pmatrix} W^{\bf\L}\cr W_{\bf\L} \end{pmatrix} \label{dXFSU33}~,
\ee
where, in terms of the rescaled sections \eq{Zresc} and \eq{Srescalintro}, we have
\ba
W^{\bf\L}&\!\!=\!\!&2n^1 e^{\varphi+\frac{K_+}2} \left[\left({\rm Re}G_{A\eta}+\k \, \tilde\z_{A\tilde\eta}\right) m^{A\L}-\left({\rm Re}Z^A_\eta+\k \,\z^A_{\tilde\eta}\right) m_{A}^{~\L}-\k\, \tilde\eta\, m^\L\right]\nn~,\\
W_{\bf\L}&\!\!=\!\!&2n^1e^{\varphi+\frac{K_+}2} \left[\left({\rm Re}G_{A\eta}+\k \,\tilde\z_{A\tilde\eta}\right) e^A_{~\L}- \left({\rm Re}Z^A_\eta +\k \,\z^A_{\tilde\eta} e_{A\L}\right)-\k\, \tilde\eta\, e_{\L}\right]\nn~,\\
&&\label{flux1}
\ea
where $\k\equiv\frac{n^3}{n^1}$ is a constant, due to \eq{Dn}, whose value depends on the specific solution.  From the left-hand side of \eq{dXFSU33} we can immediately reconstruct  $\partial_r{\rm Im}(h\,e^{U+\frac{K_+}2}\hat\Phi^0_+)$, while from the right-hand side \eq{flux1} we obtain
\be
\partial_r{\rm Im}(h\,e^{U+\frac{K_+}2}\hat\Phi^0_+)=-n^1\, e^{\varphi+\frac{K_+}2+(1-p)U}\left[ \mathcal{D} {\rm Re}\hat\Phi^0_-+\k\, \hat F_{\tilde\eta}\right]~,
\ee
where $\mathcal{D}$ is the generalised covariant derivative defined in \rf{defD}. The manipulation of \eq{attq0} is more complicated and it is less straightforward to put it into the form of a flow equation for ${\rm Im}\Phi_-$, as explained in appendix \ref{DWmanipul}. In order to get rid of the terms that do not recombine into a partial derivative,  we have  to impose $W^2=0$, that is
\be 
\langle \mathcal{D} {\rm Im}\hat\Phi^0_-,\,\hat\Phi^0_+\rangle=0\, .
\ee  
Note that this does not necessarily imply the stronger condition $\mathcal{D} {\rm Im}\hat\Phi^0_-=0$. After some manipulation we then find that we can rewrite \eq{attq0} as
\be
\partial_r\begin{pmatrix} {\rm Im}(e^{(1-\l)U+\frac{K_+}2}Z^A_{\eta})\cr{\rm Im} (e^{(1-\l)U+\frac{K_+}2}G_{\eta\, A}) \end{pmatrix} =-\frac12e^{(1-\l-p)U}\begin{pmatrix} \mathcal{E}^A\cr \mathcal{E}_A \end{pmatrix} ~,\label{dZWSU33}
\ee
where we have defined $\lambda\equiv\frac{\k^2}{1+\k^2}$, and

\be
\begin{pmatrix} \mathcal{E}^A\cr\mathcal{E}_A \end{pmatrix} =2n^1\,e^{\varphi+\frac{K_+}2} \begin{pmatrix}  {\rm Re}(hF)_{\bf\Lambda}\, m^{A{\bf\L}}- {\rm Re}(hX^{\bf \Lambda})\, e^A_{\bf\L} \cr  {\rm Re}(hF)_{\bf \Lambda}\, m_{A}^{\bf\L}- {\rm Re}(hX^{\bf \Lambda})\, e_{A\bf\L} \end{pmatrix} ~.\label{WC1}
\ee
From the left-hand side of \eq{dZWSU33} we can easily reconstruct  $\partial_r{\rm Im}(e^{(1-\lambda)U+\frac{K_+}2}\hat\Phi^0_-)$, while from the right-hand side \eq{WC1} we obtain
\be 
\partial_r{\rm Im}(e^{(1-\l)U+\frac{K_+}2}\hat\Phi^0_-)=n^1\, e^{\varphi+\frac{K_+}2+(1-\l-p)U}\mathcal{D}{\rm Re}(h\hat\Phi^0_+)~.
\ee
From the integration of \eq{dXFSU33}  and \eq{dZWSU33} we also obtain
\be
\mathcal{D}{\rm Im}(h\hat\Phi^0_+)=\mathcal{D}{\rm Im}\hat\Phi^0_-=0~.
\ee
Putting this together we find that the generalised Hitchin flow equations are given by
\ba && \label{H1}\textstyle{\frac1{n^1}}e^{-\varphi+pU}\partial_r {\rm Im} (e^{(1-\lambda)U+\frac{K_+}2}\hat\Phi^0_-)=e^{(1-\lambda)U+\frac{K_+}2}\,\mathcal{D}{\rm Re}(h\hat\Phi^0_+)~,\\
&& \label{H2}\textstyle{\frac1{n^1}}e^{-\varphi+pU}\partial_r{\rm Im}(e^{U+\frac{K_+}2}h\hat\Phi^0_+)=-e^{U+\frac{K_+}2}\left[\mathcal{D}{\rm Re}\hat\Phi^0_- +\k\,\hat F_{\tilde\eta}\right]~,\\
&& \label{H3}\mathcal{D}{\rm Im}\hat\Phi^0_-=0~,\\
&&\label{H4} \mathcal{D}{\rm Im}(h\hat\Phi^0_+)=0~.
\ea
Note that for vanishing RR fields and RR fluxes $\k=\l=0$.

In order to prepare for the comparison with the ten-dimensional result, we will rewrite \eq{H1}-\eq{H4} in a more convenient way. First, we can make use of the definitions \eq{Frescintro} and \eq{etatilde}, and use the dilaton relation \eq{10Adil} to substitute for $K_+$. Then we can use the expression for the four-dimensional dilaton that is valid for domain wall configurations $\l U=-(\varphi+U)$ (see \rf{4dimdil}) to obtain  
\ba &&\!\!\!\!  \label{H1D}\textstyle{\frac1{n^1}}e^{-\varphi+pU-2(U+\varphi)}\partial_r {\rm Im} (e^{2(U+\varphi)} e^{-\phi}\hat\Phi^0_-)=\mathcal{D}{\rm Re}(e^{-\phi}h\hat\Phi^0_+)~,\\
&& \!\!\!\! \label{H2D}\textstyle{\frac1{n^1}}e^{-\varphi+pU-(U+\varphi)}\partial_r{\rm Im}(e^{U+\varphi} e^{-\phi}h\hat\Phi^0_+)=-\mathcal{D}{\rm Re}(e^{-\phi}\hat\Phi^0_-) -\k\sqrt2\,\hat F,\\
&& \!\!\!\! \label{H3D}\mathcal{D}{\rm Im}(e^{-\phi}\hat\Phi^0_-)=0~,\\
&&\!\!\!\! \label{H4D} \mathcal{D}{\rm Im}(e^{-\phi}h\hat\Phi^0_+)=0~.
\ea



\subsection{Orientifold projection}\label{op}

In the previous subsection we considered a supersymmetric domain wall solution in $\cN=2$ supergravity preserving one half of the original supersymmetry \eq{project}. One can interpret such a solution as originating from the compactification of an appropriate ten-dimensional brane configuration filling three out of the four uncompactified spacetime directions, as we shall describe further in sections \ref{s3} and \ref{s4}. For consistency, such compactifications often require the introduction of orientifold planes which may produce a further reduction of supersymmetry. Here, we shall discuss how the flow equations that we have derived above get modified by such an orientifold-type projection.

In \cite{Andrianopoli:2001gm} and \cite{D'Auria:2005yg} it was shown how one can perform a consistent truncation of gauged $\cN=2$ supergravity with scalar and scalar-tensor multiplets, respectively, such that the $\cN=2$ multiplets are rearranged into $\cN=1$ multiplets. The identification of the fields to be truncated out involves considering the consistency of the fermionic supersymmetry transformation laws \eq{ferm}  when a linear combination of the two gravitini is set to zero\footnote{In this section we are using the notation of \cite{Cassani:2007pq}.}
\be
q^{{\hat{A}}\dagger}\p_{{\hat{A}}\m}=0~,\label{nograv}
\ee  
with the independent combination
\be
\psi_{\m +}=p^{{\hat{A}}\dagger}\p_{\m {\hat{A}}}~,
\ee
being identified with the $\cN=1$ gravitino. The projectors $p_{\hat{A}}$ and $q_{\hat{A}}$ satisfy
\be 
p^{{\hat{A}}\dagger}p_{\hat{A}}=q^{{\hat{A}}\dagger}q_{\hat{A}}=1,\quad\quad p^{{\hat{A}}\dagger}q_{\hat{A}}=q^{{\hat{A}}\dagger}p_{\hat{A}}=0\label{N=1proj}~.
\ee
Correspondingly the supersymmetry parameter combination 
\be 
\z_+=p^{{\hat{A}}\dagger}\ve_{\hat{A}}\label{e1}~,
\ee
generates $\cN=1$ supersymmetry, while the orthogonal combination
\be
\l_+= q^{{\hat{A}}\dagger}\ve_{\hat{A}}\label{e2}~,
\ee
should not appear in the $\cN=1$ supersymmetry transformation laws. It is not difficult to realise \cite{Andrianopoli:2001gm} that the only way to achieve this without spoiling the $\cN=1$ supermultiplet structure  is to set one of the two supersymmetry generators to zero $\l_+=0$, as one might expect. Consistency of \eq{nograv} then requires that the corresponding gravitino shift is set to zero
\be 
W_\perp\equiv -2i\,q^{{\hat{A}}\dagger}S_{{\hat{A}}{\hat{B}}}\,p^{*{\hat{B}}}=0~.\label{noWperp}
\ee
Solving the condition \eq{noWperp} determines which scalar fields are truncated out by the  projection and thus how the domain wall solution gets modified. The truncation of the other fermionic transformation laws provides conditions for the consistent reduction of the scalar manifolds \cite{Andrianopoli:2001gm}. We are not going to make any assumption about the preserved supersymmetry. Rather, as we did in the previous section, we will derive the BPS projectors from the supersymmetry conditions \eq{psi}-\eq{zi}, this time imposing \eq{noWperp}.  It is convenient to work from the outset in terms of the spinors \eq{e1} and \eq{e2}, for which the BPS projector \eq{project}
\be \ve_{\hat{A}}=\mp 2i\hb\,\g^3\ n_{{\hat{A}}{\hat{B}}}\ve^{\hat{B}}\label{N=2BPS}~,\ee
gives rise to
\ba
&&\z_+=\pm\hb\g_3\left[ n_{\parallel}\,\z_-+n_{\perp}\l_-\right]~,\nn\\\nn\\
&&\l_+=\pm\hb\g_3\left[n^{\dagger}_{\parallel}\l_-+n_{\perp}\z_-\right]\label{projnnn}~,
\ea
where
\be n_{\parallel}\equiv -2i\,p^{{\hat{A}}\dagger}n_{{\hat{A}}{\hat{B}}}\,p^{*{\hat{B}}},\quad\quad n_\perp\equiv -2i\,q^{{\hat{A}}\dagger}n_{{\hat{A}}{\hat{B}}}\,p^{*{\hat{B}}}~.\ee
Note that, according to \eq{normphase}, 
\be 
W_{\parallel}\equiv Wn_{\parallel}~,\quad\quad\quad W_\perp\equiv W n_\perp~.
\ee
It is natural to expect that on implementing the truncation, i.e. imposing condition \eq{noWperp}
\be 
W_\perp= n_\perp=0\label{Wperp=0}~,
\ee 
we will obtain two copies of an $\cN=1$ $\textstyle{\frac12}$ BPS condition:
\ba
&& \z_+=\pm\hb n_{\parallel}\g_3\,\z_-\label{projn}~,\\\nn\\
&& \l_+=\pm\hb n^{\dagger}_{\parallel}\g_3\,\l_-~.\label{projn*}
\ea 
In fact, let us consider again the component $\d\p_\a=0$ of equation  \eq{psi}, this time imposing $n_{\perp}=0$, according to the truncation condition \eq{noWperp}. We then find
\ba
&& U^\prime e^{pU}\g_3\z_+=n_{\parallel}W\z_-\label{psialpha1}~,\\\nn\\
&& U^\prime e^{pU}\g_3\l_+=n^{\dagger}_{\parallel}W\l_-\label{psialpha2}~.
\ea 
The consistency of these expressions again gives \eq{Uprime}. If we now insert \eq{Uprime} back into \eq{psialpha1} and \eq{psialpha2} we obtain, as expected, the projectors \eq{projn} and \eq{projn*}. Continuing as before, we can use $\d\p_{r{\hat{A}}}=0$ to obtain  
\ba
n_\parallel\hb\,\partial_r(n^{\dagger}_\parallel h)&=&\pm i e^{-pU}{\rm Im}(hW)~,\\
n^{\dagger}_\parallel\hb\,\partial_r(n_\parallel h)&=&\pm i e^{-pU}{\rm Im}(hW)~,
\ea
together with \eq{1ell} and \eq{Uprime2}. From this we can deduce that $\partial_r h=0$ and find the analogue of \eq{Dn}
\be
\partial_r n_\parallel=0~.
\ee
Using \eq{projn} and \eq{projn*} in \eq{la} and \eq{zi} we can proceed as before and obtain equations \eq{attz0} and \eq{attq0}, taking into account that due to the condition \eq{noWperp} the flow equations \eq{attz0} and \eq{attq0} are restricted to the set of scalars that survive the projection.  
From \eq{psi} we again obtain \eq{4drsoln}, which in the new basis reads
\ba
\z_+&=&e^{\frac12 U}\z_+^0\label{KSz}~,\\
\l_+&=&e^{\frac12 U}\l_+^0\label{KSl}~,
\ea 
where $\z_+^0$ and $\l_+^0$ are constant spinors subject to \eq{projn} and \eq{projn*}.


As shown in \cite{D'Auria:2005yg} one can obtain three classes of projection of the $\cN=2$ theory. Let us define
\be p_{\hat{A}}=\begin{pmatrix} a^*\cr b \end{pmatrix} ~,\ee
where the parameters $a$ and $b$ are complex constants satisfying (via \eq{N=1proj}) $|a|^2+|b|^2=1$. For orientifold projections $a$ and $b$ further satisfy  $|a|^2-|b|^2=0$ and are related to the phase $\theta$ in \eq{Ccomp} by  $e^{i\theta}=2|a|^2b/a^*$. The three types of projection are the ``Heterotic'' projection $\left(a=1,\,b=0\right)$, which projects out all the RR fields, and two orientifold projections $\left(a=\frac1{\sqrt 2},\,b=-\frac{i}{\sqrt2}\right)$, $\left(a=\frac1{\sqrt 2},\,b=\frac{1}{\sqrt2}\right)$. For these different values of $a$ and $b$ we obtain 
\ba
\label{HH} (H)\quad\quad\quad a=1,\ b=0~,\quad&&\quad n_\perp=-n^3=0,\quad\quad n_\parallel=(n^1+in^2)~,\nn\\\nn\\
\label{B}(B)\quad a=\textstyle{\frac1{\sqrt 2},\ b=-\frac{i}{\sqrt2}}~,\quad&&\quad n_\perp=n^2=0,\quad\quad \ \ \, n_\parallel=(n^1+in^3)~, \nn\\\nn\\
\label{C}(C)\quad a=\textstyle{\frac1{\sqrt 2},\ b=-\frac{1}{\sqrt2}}~,\quad&& \quad n_\perp=n^1=0,\quad\quad\ \ \, n_\parallel=(n^3+in^2)~.
\nn
\ea
It is straightforward to see that $n_{||}$ is pure phase using \rf{normphase} and \rf{Wperp=0}. The condition \eq{Wperp=0} corresponds to $W^3=0$, $W^2=0$, $W^1=0$ in the three cases $H$, $B$ and $C$, respectively. This is enough to identify the reduction of the scalar sector and, hence, the corresponding truncation.   

Note that when manipulating  \eq{attq0} we found that we could only write a flow equation for ${\rm Im}\hat\Phi^0_-$ when $W^2=0$. This meant that a solution corresponding to the geometry described by \eq{H1}-\eq{H4} naturally splits $\cN=2$ supersymmetry into two copies of $\cN=1$ via the choice $B$. This suggests that the ten-dimensional configurations that give rise to such domain wall solutions feature orientifold planes with $\theta=\pm\frac\pi2$ \footnote{The $\pm$ sign is due to the ${\rm SU}(2)$ symmetry of the problem that allows us to exchange $\ve_1$ with $\ve_2$ by acting with $\s^1$ \cite{D'Auria:2005yg}. In our case, this would just correspond to a different choice of $a$ and $b$.}.

Finally, we shall briefly discuss the truncation $H$, since it is likely to correspond to an $SU(3)\times SU(3)$  structure compactification of the Heterotic theory at the zeroth order in $\a^\prime$. As in \cite{D'Auria:2005yg} we can see that the constraint $W_\perp=W^3=0$ implies that all the RR scalar fields and fluxes are identically zero 
\be\z^A=\tilde\z_A=e_{\bf\L}=m^{\bf\L}=0~,\ee
and that the $\cN=1$ K\"ahler-Hodge manifold is the product of the complex structure deformations and the K\"ahler class deformations. As there is no projection acting on the Calabi-Yau manifold, the definition of $\hat\Phi_{\pm}$ is the same as in the $\cN=2$ case \eq{p1}-\eq{p2}. Finally, in the absence of the RR part, $n^1=1$ and $\k=\l=0$, the equations \eq{H1D}-\eq{H4D} read
\ba 
&& d_H{\rm Re}(e^{-\phi}\,h\hat\Phi^0_+)=\partial_y {\rm Im}(e^{-\phi}\,\hat\Phi^0_-)~,\\
&& d_H{\rm Re}(e^{-\phi}\,\hat\Phi^0_-)=-\partial_y {\rm Im}(e^{-\phi}\,h\hat\Phi^0_+)~,\\
&&d_H{\rm Im}(e^{-\phi}\,\hat\Phi^0_-)=0~,\\
&&d_H{\rm Im}(e^{-\phi}\,h\hat\Phi^0_+)=0~,
\ea
where we have defined a new transverse coordinate
\be 
\partial_y\equiv e^{(1+p)U}\partial_r~.
\ee
From the ten-dimensional perspective it is known that there are subtleties in the analysis of pure spinor equations for heterotic compactifications (see \cite{Andriot:2009fp} for a recent discussion), and so we shall leave a detailed discussion of this case for later work.


\section{Domain walls in 4D from 10D supergravity}\label{s3}

We will now turn to the derivation of the flow equations from the ten-dimensional perspective. We want to have a ten-dimensional type II supergravity configuration which gives rise to a domain wall in the effective, four-dimensional description. Therefore, we shall consider an ansatz for a spacetime of the form $M^{1,2}\times_w\mathbb{R}\times_w \hat{Y}$, where $\hat{Y}$ is a SU(3)$\times$SU(3) structure manifold and the products are warped. As we shall make use of the democratic formalism \cite{bkorp}, it is most convenient to work in the string frame. We will take the following general form for the metric:
\begin{equation}
ds^2 = e^{2A(y,r)}\left( e^{2V(r)}\eta_{\a\b}dx^\a dx^\b + e^{2G(r)}dr^2 \right) + g_{mn}(r,y)dy^m dy^n~,\label{metric}
\end{equation}
where now $\a,\b=0,1,2$ label the domain wall worldvolume directions, which are flat, and  $m,n=1,\ldots 6$ label directions on $\hat{Y}$. $A(y,r)$ is called the warp factor and we want the term in brackets to describe a supersymmetric domain wall solution in four dimensions. We shall allow the ten-dimensional dilaton $\phi$ to depend on the transverse and internal coordinates $\phi = \phi(y,r)$, as is appropriate for a domain wall configuration.

Let us introduce the modified RR field strengths
\bea
F_{(n+1)}=dC_{(n)}+H\wedge C_{(n-2)}\ ,
\eea
where $dC_{(n)}$ are the standard RR field strengths\footnote{We are essentially following the conventions of \cite{gmpt1,gmpt2}, up to some differences consisting in a sign for $H$ in type IIB  and the sign change $C_{(2n+1)}\rightarrow (-)^nC_{(2n+1)}$ in type IIA.}. The most general RR flux decomposition respecting the domain wall symmetry is
\begin{equation}
F^{(10)}_{n} = \mathrm{vol_{dw}}\wedge f^{||}_{n-3} + dr\wedge f^{\bot}_{n-1} + \hat{F}_{n} + \mathrm{vol_{4}}\wedge \tilde{F}_{n-4}~, \label{RR}
\end{equation}
where $\mathrm{vol_{dw}}$ and $\mathrm{vol_{4}}$ denote the obvious volume forms on the domain wall $\mathbb{R}^{1,2}$ and the total four-dimensional external space $\mathbb{R}^{1,2}\times\mathbb{R}_r$, respectively (both viewed from ten dimensions). All $f$'s and $F$'s are forms on the internal manifold $\hat{Y}$. For domain walls the $\hat{F}_p$ and $\tilde{F}_p$ are pure internal and external $p$-form fluxes, respectively. In type IIA, the index $n$ runs over $0, 2, 4, 6, 8, 10$ while in type IIB $n$ runs over $1, 3, 5, 7, 9$. From now on, we shall set $f^{||} = 0 = f^{\bot}$, in agreement with the choice of section \ref{s2} where the tensor fields were set to zero. In four dimensions such tensor fields would correspond to fluxes on the domain wall worldvolume and are not considered in \cite{Grana:2006hr}, to which we want to make contact.

The RR fluxes described above contain both field strengths and their duals, so we must impose the self-duality relations
\begin{equation}\label{sd}
F^{(10)}_{(n)}=(-)^{\frac{(n-1)(n-2)}{2}}\star_{10} F^{(10)}_{(10-n)}~,
\end{equation}
between the lower and higher rank  field strengths. Unless explicitly stated otherwise, we shall always make use of the self-duality relations to write the RR fields entirely in terms of $\hat{F}$. 

The NS flux is decomposed in a similar manner as follows:
\begin{equation}\label{NS}
H^{(10)} = H_3 + dr\wedge b'_2~,
\end{equation}
where $H_3$ and $b_2(y,r)$ are forms on $\hat{Y}$, and $'$ denotes a transverse derivative $\pt/\pt_r$. Once again, for simplicity we shall only consider the $b_2 = 0$ case here.

\subsection{Analysing the supersymmetry variations}

The type II gravitino and dilatino supersymmetry variations in string frame are
\begin{eqnarray}
\delta\psi_M &=& (D_M \pm \frac{1}{4} H_M \cP)\epsilon +
\frac{e^{\phi}}{16} \sum_n \slash{\hat{F}}_{2n} \Gamma_M\cP_n{}\epsilon~, \\
\delta\lambda & = & (\Gamma^M{\partial}_M\phi \pm \frac{1}{2} \slash{H}\cP)\epsilon
+
\frac{e^{\phi}}{16} \sum_n \Gamma^M\slash{\hat{F}}_{2n}\Gamma_M \cP_n{}\epsilon ~,
\end{eqnarray}
where one chooses the upper sign for IIA and the lower sign for IIB. Capital Latin letters run over all ten directions. These expressions are written in the democratic formalism \cite{bkorp} with all spinor indices suppressed; so, for instance, $\epsilon = (\epsilon^1, \epsilon^2)$ is a doublet of ten-dimensional Majorana-Weyl spinors. The $\cP$ matrices act on these doublets as 
$\cP = \Gamma_{11}$ and $\cP_n = \Gamma_{11}^n\sigma^1$ in type IIA, and as $\cP = -\sigma^3$, $\cP_{n} = \sigma^1$ for $(n/2+1/2)$ even and $\cP_n = i\sigma^2$ for $(n/2+1/2)$ odd in type IIB. We will also make use of the modified dilatino variation \cite{gmpt2}, 
\begin{eqnarray}
\d \L = \G^M \psi_M - \d \l = 0 \, \label{mod0}~,
\end{eqnarray}
in which all RR terms cancel. 

We begin by substituting our ansatz for the metric \rf{metric}, the RR fields \rf{RR} and the NS field \rf{NS} into the supersymmetry variations,
\begin{eqnarray}
\delta\psi^1_\a &=& \half \G_\a \slash{\pt}A \e^1 + \half e^{A-G}(V'+A')\G_{\ul{\a r}}\e^1 - \frac{e^\phi}{8}\slash{\hat{F}} \Gamma_\a\epsilon^2 \label{wv0} =0 ~,\\
\delta\psi^1_r &=& \pt_r\e^1 + \half \G_r \slash{\pt}A \e^1 - \frac{e^\phi}{8}\slash{\hat{F}} \Gamma_r \epsilon^2 \label{radial0}  =0 ~,\\ 
\delta\psi^1_m &=& \left(D_m + \frac{1}{4} H_m \right)\epsilon^1 - \frac{1}{4} \G^{rn} g'_{mn} \e^1 - \frac{e^\phi}{8}\slash{\hat{F}} \Gamma_m \epsilon^2 = 0\label{int0}~, \\
\delta\Lambda &=& \left( \slash{D} - \slash{\partial}\phi + \frac{1}{4} \slash{H} + 2\slash{\partial}A \right)\e^1 + (2V'+2A'-\phi')e^{-(A+G)}\G_{\ul{r}}\e^1 \nn \\ &~&+ ~~ \frac{1}{4} g^{mn} g'_{mn} \G^r \e^1 = 0 \label{md0}~,
\end{eqnarray} 
where we have made the standard ansatz that $\epsilon$ is independent of the worldvolume coordinates $\epsilon = \epsilon(y,r)$. Underlined indices are flat tangent space indices. From now on it should be understood that slashed quantities are purely internal e.g. $\slash{\partial}\phi \equiv \Gamma^m \partial_m\phi$. The $\epsilon^2$ variations are found from the expressions above by taking the map 
\begin{eqnarray}
\slash{H} \rightarrow -\slash{H},~~~~\slash{\hat{F}} \rightarrow + \slash{\hat{F}}^\dagger \label{map}~,
\end{eqnarray}
and interchanging $\epsilon^1$ and $\epsilon^2$.

The transverse component of the gravitino variation $\delta\psi^1_r$ plays an important role in what follows. We want to manipulate this component so that we can determine the transverse dependence of the spinor parameter $\epsilon$. By comparing with the modified dilatino variation, it is straightforward to see that we can use the worldvolume component of the gravitino variation to simplify the transverse component. Specifically, we calculate $\G^\a \d\psi^1_\a$ and use the result to substitute for the $\slash{\pt}A$ and RR terms in $\d\psi^1_r=0$ to find
\begin{equation}
\delta\psi^1_r = \pt_r\e^1 -  \half (A' + V')\e^1 = 0~. \label{radial1}
\end{equation}
This is easily solved by a typical domain wall ansatz, factoring out the transverse dependence\footnote{While \rf{rsoln} appears to be the same as \rf{4drsoln}, the reader is reminded the two expressions are written in different frames. We shall carry out a comparison in section \ref{s4}.}:
\begin{equation}
\epsilon(r, y^m) =  e^{\half(A+ V)}\epsilon_0(y^m)~.\label{rsoln}
\end{equation}
If we then substitute this back into \rf{radial0} we find 
\begin{equation}
\delta\psi^1_r =  \half (A' + V')\epsilon^1   + \half \G_r \slash{\pt}A \e^1 - \frac{e^{\phi}}{8}\slash{\hat{F}} \Gamma_r \epsilon^2 =0~ \label{radial2}~,
\end{equation}
which is the same as $\d\psi^1_\a=0$. We should now decompose the ten-dimensional quantities appearing here into four- and six-dimensional components. This requires us to make an ansatz for the spinor decomposition and for the projection condition enforced by the domain wall.

\subsection{$\cN=1$ spinor ansatz and the BPS projection condition}
  
Throughout the rest of this section will shall focus on type IIA supergravity, as the type IIB case proceeds analogously.  For type IIA backgrounds the supersymmetry parameter is a ten-dimensional Majorana spinor $\epsilon$ that can be split into two Majorana-Weyl spinors of opposite chirality. In order to harness the compact pure spinor notation found in the equations describing $\cN=1$ Minkowski and AdS vacua \cite{gmpt2}, we shall consider backgrounds which initially preserve at least four supercharges before any domain wall projection conditions are applied. The Killing spinors can be decomposed as
\bea\label{spinors} 
\epsilon^1_0(y) &=&\zeta_+\otimes
\eta^{(1)}_{+}(y)+\zeta_-\otimes \eta^{(1)}_{-}(y)\ ,\cr
\epsilon^2_0(y) &=&\zeta_+\otimes
\eta^{(2)}_{-}(y)+\zeta_-\otimes \eta^{(2)}_{+}(y)\ , 
\eea 
where $\zeta_{+} = (\zeta_{-})^*$ is a generic constant four-dimensional spinor of positive chirality, while the $\eta^{(a)}_+=(\eta^{(a)}_-)^*$ are two particular six-dimensional commuting spinors of positive chirality that characterise the solution. In the usual abuse of notation, we use the subscripts $\pm$ to denote both four- and six-dimensional chirality. The norms of the internal spinors are defined as
\bea
||\eta^{(1)}||^2=|a|^2~,\quad\quad ||\eta^{(2)}||^2=|b|^2\ . \label{norms}
\eea

We would like to find a domain wall solution preserving at least $1/2$ of the $\cN=1$ supersymmetry in four dimensions i.e. 2 supercharges. Therefore, we should expect the four four-dimensional supersymmetries to be related by a projection condition. Motivated by the probe Dp-brane supersymmetry projection for general $\cN=1$ backgrounds \cite{ms,Koerber:2005qi}, we make the following ansatz for the domain wall projection condition:
\begin{equation}\label{proj1}
\gamma_{\ul{0\ldots 2}}\zeta_+=\alpha^{-1}
\zeta_{-} \,~~ \Longrightarrow ~~ \gamma_{\ul{r}}\zeta_+= i \alpha^{-1} \zeta_{-}~.
\end{equation}
Consistency of this ansatz requires that $\a$ is pure phase $\a^{-1} = \a^*$. Following \cite{hmt}, it would be straightforward to extend our spinor ansatz \rf{spinors} and the projection condition \rf{proj1} to allow for a greater amount of supersymmetry. The projection condition \rf{proj1} is a physically well-motivated choice as it agrees with the supersymmetry constraints on $\half$-supersymmetric domain walls in $\cN=1$ supergravity in four dimensions \cite{Cvetic:1996vr} and with \rf{projn}. 

It is natural to ask whether there is any relation between $\a$, $a$ and $b$. For probe D-branes in Minkowski or AdS background certain relations have been found for specific examples in \cite{Aharony:2008wz}. However, as this appears to be an example-dependent feature we shall not impose any further relations between the phase and the complex coefficients here. In general, we shall only require that the domain wall should preserve two of the four four-dimensional supercharges, corresponding to $\cN=1$ supersymmetry on its worldvolume. From the ten-dimensional perspective this amounts to a preservation of $1/16$th supersymmetry. We shall not pursue this point any further, but refer the reader to \cite{Caviezel:2008ik} for a discussion of $1/16$th-supersymmetric type II intersecting brane configurations that give rise to domain walls after dimensional reduction on cosets and nilmanifolds.

\subsection{Pure spinors and Hitchin flows for domain wall vacua}

Using our ansatz for the projection condition \rf{proj1}, along with the $4+6$ decomposition of the spinors \rf{spinors}, we can now rewrite our supersymmetry variations in terms of pure spinors. We shall give the result here and refer to our appendix \ref{psd} and appendix A of \cite{gmpt3} for further details of the calculations. We construct the normalised pure spinors from bispinors as follows:
\begin{equation}
\sph_\pm = -\frac{8 i}{|a|^2} \eta^{(1)}_+\otimes\eta^{(2)\dagger}_{\pm}~. \label{nps}
\end{equation}
By virtue of \rf{rsoln} and \rf{spinors}, the bispinors $\eta^{(1)}_+\otimes\eta^{(2)\dagger}_{\pm}$ are independent of the transverse coordinate. However, using the Clifford map \rf{cmap} we see that the related differential polyforms have $r$-dependence through the vielbein. As is usual in the literature, we shall drop the slash notation which distinguishes the polyform and bispinor.  We would now like to rewrite the supersymmetry conditions derived above in terms of differential constraints on $\Phi_\pm$ . After a lengthy calculation we find\footnote{These equations have been derived independently in \cite{hlmt}.} 
\begin{eqnarray}
d_H \left[e^{2A - \phi} \mathrm{Im} \Phi_- \right] \!\! &=&\!\! 0 \label{ps1} ~,\\
d_H \left[e^{4A - \phi} \mathrm{Re} \Phi_- \right] \!\!&=&\!\! e^{4A} \tilde{F} - e^{-3V-G} \mathrm{Im} \left( \a^* \pt_r \left[ e^{3A +3V -\phi} \Phi_+ \right]\right), \label{ps2} \\
d_H \left[e^{3A-\phi} \mathrm{Im} \left(\a^* \Phi_+\right) \right] \!\!&=&\!\! 0  \label{ps3}~,\\
d_H \left[e^{3A-\phi}  \mathrm{Re} \left(\a^* \Phi_+\right) \right] \!\!&=&\!\! e^{-2V-G} ~ \pt_r \mathrm{Im} \left[ e^{2A+2V-\phi} \Phi_- \right] ~,\label{ps4}
\end{eqnarray}
where $d_H \equiv d + H\wedge$ is the twisted exterior derivative on the six-dimensional manifold $\hat{Y}$. The result for type IIB is found by taking the map \rf{map}. In deriving these expressions we have made use of the following additional constraint derived from $\d\psi_m$ 
\bea
\label{backsusy2} 
d|a|^2=|b|^2dA~,\quad\quad d|b|^2=|a|^2dA\ . 
\eea
Note that we have used $\tilde{F} = \star ~{\sigma (\hat{F})}$, where $\sigma$ is an involution which reverses the order of indices on a form, to rewrite the RR fluxes $\hat{F}$ in terms of their duals ${\tilde{F}}$. This makes it straightforward to verify that the above equations agree with those presented in \cite{gmpt3,ten2four,adsbranes}, in the limit where the four-dimensional component of the spacetime becomes $\mathrm{Minkowski}$ or $AdS$.

In general, one finds that the right-hand side of \rf{ps1} is non-vanishing and proportional to $(|a|^2-|b|^2)\hat{F}$. As described in section \ref{op}, orientifold projections enforce $|a|^2=|b|^2$, and in \cite{ms} it was shown that the same condition is necessary for an $\cN=1$ Minkowski spacetime to admit supersymmetric probe D-branes. For anti-de Sitter spacetimes, one can show that $|a|^2=|b|^2$ is a consistency condition of the background itself \cite{gmpt3}, independent of the probe D-brane argument. When the warp factor $A$ is independent of the transverse direction $r$, it is straightforward to check that the same holds true of the domain wall background \rf{metric}. 

We can also rewrite the four external components of the gravitino variations, \rf{wv0} and \rf{radial0}, in terms of the pure spinors. We have already shown that the transverse dependence of the spinor parameter $\e$ can be factored out \rf{rsoln}, leaving us with just one equation \rf{radial2}, which was used in the derivation of the (\ref{ps1}-\ref{ps4}). Equation \rf{radial2} is a transverse flow equation for the metric data $A$ and $V$, with a potential given in terms of $F$ and $dA$. Following \cite{hmt,Grana:2006hr}, we will rewrite the potential using the Mukai pairing on the internal manifold \rf{Mukai}. Multiplying \rf{radial2} on the right by $\eta^{(2)\dagger}_{+}$ and taking the spinor trace we find
\begin{equation}
(A'+V') = - \frac{i \a e^\phi}{4} \frac{\langle \hat{F}, \overline{\Phi}_{\pm} \rangle}{\langle \Phi_{-}, \overline{\Phi}_{-} \rangle}~, \label{auflow}
\end{equation}
and we see that the $dA$ contribution drops out by virtue of the compatibility condition \rf{comp}. In the following section we will show how equations (\ref{ps1}-\ref{ps4}) and \rf{auflow} agree with those derived by analysing the flow of the vector and hypermultiplet scalar fields in four dimensions.

\subsubsection*{Hitchin flow}

We have shown that the set of equations \rf{ps1}-\rf{ps4} defined in terms of pure spinors on $\hat{Y}$, along with \rf{backsusy2}, describe the necessary conditions for a domain wall solution in four dimensions preserving at least 2 supercharges. Following the literature discussing the cases with vanishing fluxes \cite{Witt} (see also section \ref{s4.1}), it is interesting to ask whether there is a 7-dimensional interpretation of these results, where the manifold $\hat{Y}$ is fibred over an interval given by the direction transverse to the domain wall. In fact, following the discussion of the $AdS_4$ backgrounds in \cite{adsbranes}, it is relatively straightforward to show that this set of equations can be rewritten in terms of a generalised $G_2$ structure defined by \cite{pk}\footnote{We thank Paul Koerber for useful discussions on this point.}
\begin{eqnarray}
 \rho &=& e^{A+G} dr \wedge \mathrm{Re} \Phi_- \mp \mathrm{Im} \left(\a^* \Phi_+\right) \, ~,\label{r}\\
 \hat \rho &=&  - \mathrm{Im} \Phi_- \mp e^{A+G} dr \wedge \mathrm{Re} \left(\a^* \Phi_+\right)~,\label{rh}
\end{eqnarray}
where $\rho$ and $\hat\rho$ are related by the generalised Hodge star in seven dimensions. It is then possible to rewrite the six-dimensional pure spinor equations in terms of the seven-dimensional quantities as 
\begin{eqnarray}
 \hat d_H \left[e^{2(A+V)-\phi}\hat\rho \right] &=& 0 ~, \\
\hat d_H \left[e^{3(A+V)-\phi} \rho \right] &=&  - e^{4A +3V + G} dr \wedge \tilde{F}~,
\end{eqnarray}
where now $\hat d_H = d_H + dr\pt_r$ is the twisted exterior derivative in seven dimensions. These equations define a generalised $G_2$ structure $\rho$ which is integrable with respect to $H$ and $F$, or, alternatively, an almost generalised $G_2$ structure with $F$ providing the obstruction to integrability \cite{Jeschek,adsbranes}. This implies that our set of equations \rf{ps1}-\rf{ps4} are a form of \textit{generalised Hitchin flow} equations describing the embedding of the $SU(3)\times SU(3)$ structure manifold $\hat{Y}$ into a generalised $G_2$ manifold, where one now has $G_2\times G_2$ structure defined on the formal sum of the tangent and cotangent bundles of the  seven-manifold. Written in seven-dimensional notation, these equations are the same as their Minkowski \cite{Jeschek} and $AdS$ counterparts \cite{adsbranes}. One only notices a difference on decomposing $\rho$ and $\hat\rho$ into forms on $\hat{Y}$, when the more general $r$-dependence of the pure spinors on a domain wall background becomes apparent. In Hitchin's original language \cite{HitchinHF}, the case with $F=0$ was called a strongly integrable generalised $G_2$ structure. When the flux contribution $dr\wedge \tilde{F}$ is identified as being proportional to $\rho$, the flow equations give Hitchin's definition of a weakly integrable generalised $G_2$ structure.

\section{Comparing the flow equations}\label{s4}

In this section we shall show that our domain wall vacua equations derived in ten and four dimensions are in agreement. By directly studying the supersymmetry variations for domain wall configurations we are essentially carrying out an on-shell (up to imposing Bianchi identities \cite{Koerber:2007hd}) test of the procedure proposed in \cite{Grana:2006hr} for $SU(3)\times SU(3)$ structure compactifications. The analogous check for $\cN=1$ Minkowski and AdS vacua has been carried out in \cite{Cassani:2007pq}, where it was shown that the pure spinor equations found in ten dimensions \cite{gmpt2} agree with those found by first carrying out an $SU(3)\times SU(3)$ structure compactification \cite{Grana:2006hr} and then looking at the conditions for maximally symmetric vacua. A key point in this check is the relation between the pure spinors in ten dimensions \rf{nps} and the Kaluza-Klein truncated pure spinors in the four-dimensional effective theory \rf{p1}-\rf{p2}. For $SU(3)\times SU(3)$ structure compactifications the truncation of the pure spinors such that appropriate special K\"ahler geometry arises in the kinetic terms of the resulting four-dimensional theory has been fully discussed in \cite{Grana:2006hr}. We have reviewed some necessary details of this in section \ref{s2} and appendix \ref{4dstr}, but for the remainder of this section it is sufficient for the reader to remember that one can make a rigourous comparison of the pure spinors.   

In order to compare our flow equations we have to pay attention to our conventions, in particular the metric signature and the choice of chirality assigned to the spinors. The supersymmetry analysis in section \ref{s2} employed the mostly-minus metric signature ($+,-,-,-$), in keeping with with previous four-dimensional supergravity conventions \cite{Louis:2006wq}, whereas the ten-dimensional analysis used the mostly-plus signature ($-,+,\cdots,+$) which is more common in the flux compactification literature \cite{Grana:2005jc}. We can rewrite our ten-dimensional expressions in the mostly-minus convention by inserting a minus sign for any explicit upper index or metric factor $g_{MN}$ and multiplying all gamma matrices $\G_M$ by $i$. Of particular importance is the projection condition \rf{proj1}, which becomes
\begin{equation}\label{proj2}
\zeta_+= \alpha^* \gamma_{\ul{r}} \zeta_{-}~.
\end{equation}
By comparing this with the four-dimensional projector \rf{projn}, we find that we should make the following identifications
\be
\zeta_+ \rightarrow (n_{\parallel}^*)^{\frac{1}{2}} \zeta_+~, \qquad \alpha = h ~.
\ee
The rescaling of spinors $\zeta_+$ can be understood as a K\"ahler transformation in four dimensions, and is equivalent to the $\mathbb{C}^*$ action on the pure spinors \rf{nps}. 

References \cite{Grana:2005ny,Cassani:2007pq} assigned negative parity to $\e^1$ and positive parity to $\e^2$, whereas we have used the opposite convention, see \rf{spinors}. Our pure spinors ${\Phi}_\pm$ are mapped to the complex conjugate of those used in \cite{Grana:2005ny,Cassani:2007pq}, denoted $\overline{\Phi^{0}_\pm}$, and we also need to send $H\rightarrow -H$. Therefore, we should work out the supersymmetry equations for $\overline{{\Phi}_\pm}$ and map them to equations in terms of $\Phi^{0}_\pm$ using\footnote{Unlike \cite{Cassani:2007pq}, we do not need to map the fluxes as we have used the uniform notation of \cite{ms} throughout.}:
\be \overline{\Phi_+} ~~\longrightarrow~~ \Phi^{0}_+ = X^{\bf\L}\o_{\bf\L}-F_{\bf\L}\o^{\bf\L}~,\ee
\be  - \overline{(\Phi_-)}~~ \longrightarrow ~~ \Phi^{0}_-= Z^A\a_A-G_A\b^A~,\ee
The pure spinors $\Phi^{0}_\pm$ are now the same as those employed in section \ref{s2}, and they have been expanded on a finite basis of forms. As we found that $|a|^2=|b|^2$ for domain wall backgrounds, the complex coefficients $a$ and $b$ in the pure spinors \rf{nps} give rise to two combinations of phases. In fact, it is only the sum of the phases that is of physical significance, and it convenient to set the phase in $\Phi_+$ to be $1$. For instance, in the $SU(3)$ structure case we then have
\bea
\Phi^0_+= e^{iJ}~,\quad\quad \Phi^0_-= e^{2 i \theta}\Omega_\eta\ ,
\eea
where now $\theta$ is identified with the phase appearing in the compensator field $C$ \rf{Ccomp}. The normalisation \rf{spinnorm} and compatibility \rf{comp} conditions for the SU(3) structure case are
\bea
J\wedge \Omega_\eta=0 \, , \qquad \frac{1}{3!}J\wedge J\wedge J=-\frac{i}{8}\Omega_\eta\wedge\bar\Omega_\eta~.
\eea
 
Let us look at how this effects the flow equations. In the reduction to four dimensions, the warp factor $A$ \eq{metric} is neglected \cite{Grana:2005ny,Grana:2006hr}; hence, in order to perform our comparison we will set $A=0$ in \rf{ps1}-\rf{ps4}:  
\begin{eqnarray}
d_H \left[e^{- \phi} \mathrm{Im} \left( \Phi^0_-\right) \right] &=& 0 \label{nps1} \\
d_H \left[e^{- \phi} \mathrm{Re} \left( \Phi^0_-\right) \right] &=& - \tilde{F} - e^{-3V-G} \pt_r \left( \mathrm{Im} \left[ e^{3V -\phi} h  \Phi^0_+ \right] \right), \nn \\  \label{nps2} \\
d_H \left[e^{-\phi} \mathrm{Im} \left(h  \Phi^0_+\right) \right] &=& 0  \label{nps3}\\
d_H \left[e^{-\phi}  \mathrm{Re} \left(h  \Phi^0_+ \right) \right] &=& e^{-2V-G} ~ \pt_r \left(\mathrm{Im} \left[ e^{2V-\phi}~  \Phi^0_- \right]\right) ~,\label{nps4}
\end{eqnarray}
where now $d_H = d - H\wedge$. It is this set of equations \rf{nps1}-\rf{nps4} that should be used to compare with the results from section \ref{s2}. To proceed further, we need a dictionary between the four-dimensional quantities $U$, $\varphi$ and their ten-dimensional counterparts $V$, $G$.  The standard relation between the ten- and four-dimensional Einstein frame metrics is
\begin{equation}
ds^2_{10_{E}} = \mathrm{vol}_{6_E}^{-1}~ds^2_{4_E}+ g_{mn}^E(r,y)dy^m dy^n~, \label{mrel}
\end{equation}
where $\mathrm{vol}_{6_E}$ is the volume of the internal manifold in ten-dimensional Einstein frame. Similarly, the dilatons $\phi$ and $\varphi$ are related by $e^{-2\varphi} = e^{-2\phi} \mathrm{vol}_{6_S}$, where now $\mathrm{vol}_{6_S}$ is the volume of the internal manifold in ten-dimensional string frame. Recall that the ten-dimensional Einstein frame metric $g_{MN}^{~\mathrm{E}}$ and string frame metric $g_{MN}^{~\mathrm{S}}$ are related by $g_{MN}^{~\mathrm{S}} = e^{\frac{\phi}{2}}g_{MN}^{~\mathrm{E}}$; therefore, the volumes are related by $\mathrm{vol}_{6_S} = e^{\frac{3\phi}{2}} \mathrm{vol}_{6_E}$. Putting this together we find the relation between the ten-dimensional string frame metric and the four-dimensional Einstein frame metric
\begin{equation}
 ds^2_{10_S} = e^{2\varphi} ds^2_{4_E} + \cdots~.
\end{equation}
We can now apply this to our metrics \rf{metric} and \rf{4dmetric} to match the parameters as follows
\begin{equation}
V = \varphi + U~, \qquad G=\varphi -pU~. \label{VG}
\end{equation}

As an initial consistency check we can compare the expressions for the Killing spinors in four dimensions. The four-dimensional component of the Killing spinor that comes from decomposing the ten-dimensional solution \rf{rsoln} naturally appears in string frame, $ \zeta =  e^\frac{V}{2}\zeta_0~$,
where $\zeta_0$ is an arbitrary constant spinor parameter. We can rescale this to Einstein frame in four dimensions using $\zeta \rightarrow e^{-\frac{\varphi}{2}}\zeta$. If we now use the matching of the metric factors \rf{VG} we find $ \zeta =  e^\frac{U}{2}\zeta_0$, which agrees with the result found from the four-dimensional approach \eq{KSz}.

Returning to the flow equations \rf{nps1}-\rf{nps4}, one can see that in order to have agreement we need to perform the following rescaling: 
\be
\Phi^0_\pm\rightarrow e^{-2V}\Phi^0_\pm,\quad\quad \tilde{F}\rightarrow \k \sqrt{2}~ e^{-2V}F ~.
\ee
The rescaling of the pure spinors is easily achieved by appropriately fixing an overall $V$-dependence of the internal metric $g_{mn}(r,y)$. However, the source of the $V$-rescaling of the RR term $F$ is unclear. We believe that it is linked to a necessary rescaling of kinetic terms in the effective action which will become clear upon a careful analysis of the dimensional reduction, and we shall pursue this point further in future work \cite{SV2}. Despite this, it is pleasing to see that we are able to find an agreement between the domain wall vacuum equations in the four- and ten-dimensional approaches.
 
Finally, let us now return to the equation derived from \rf{radial2} describing the behaviour of $V$ (see \rf{auflow}):
\begin{equation}
V'= - \frac{i \a e^\phi}{4} \frac{\langle \hat{F}, \overline{\Phi}_{\pm} \rangle}{\langle \Phi_{-}, \overline{\Phi}_{-} \rangle}~, \label{auflow2}
\end{equation}  
At first glance one might worry that this has not captured all the relevant terms in the transverse flow. From the four-dimensional perspective, we have already seen that the potential term in the transverse flow of the four-dimensional metric component $U$ is given by $W$ \rf{attu}, which contains RR flux and torsion terms due to the non-closure of the pure spinors. This discrepancy is resolved by noting that a simple splitting of the ten-dimensional Einstein frame gravitino kinetic term using $\psi_M = (\psi_\mu, \psi_m)$ does not lead to a canonical kinetic term for the four-dimensional gravitino component \cite{Grana:2005ny,Grana:2006hr}. Rather, one should consider the combination $\Psi_\mu = \psi_\mu + \frac{1}{2}\gamma_{\m}^{~m}\psi_m$. Applying this reasoning to the ten-dimensional supersymmetry transformations one finds precisely the torsion terms $\langle \Phi_{+}, d\overline{\Phi}_{-} \rangle$ that are missing from the right-hand side of \rf{auflow2}, in agreement with the four-dimensional result of section \ref{s2}. We shall not write the expression here as it is equivalent to the four-dimensional result given earlier.


\subsection{Examples}\label{s4.1}
In this section we shall discuss some particular examples of the flow equations  \eq{ps1}-\eq{ps4} with certain fluxes vanishing. Switching on the different kinds of deformations, we find different behaviours for the field $\phi$ and $K_+$, and different classes of metrics. The simplest metric is obtained when $W^3=0$, as in this case the four-dimensional dilaton is given by \eq{4dimdil}
\be 
e^\varphi=e^{-U}\label{4dimdilF=0}~,
\ee 
and, consequently, the relations \eq{VG} simplify to give
\be 
V=0~,\quad\quad G=-(1+p)U\,.\label{VGF=0}
\ee
The ten-dimensional metric then reads 
\begin{equation}
ds_{10}^2 =  \eta_{\a\b}dx^\a dx^\b+ e^{-2(1+p)U(r)}dr^2 + g_{mn}(r,y)dy^m dy^n~.
\end{equation}
Note that the factor $G$ can always be absorbed by a rescaling $dz\equiv e^{-(1+p)U(r)}dr$; thus, the resulting ten-dimensional metric is just that of $\mathbb{R}^{1,3}\times \hat{Y}$, where $\hat{Y}$ is an ${\rm SU(3)}\times SU(3)$ structure manifold
\begin{equation}\label{metricF=0}
ds_{10}^2 =ds^2_{\mathbb{R}^{1,2}}  +dz^2 + ds^2_{\hat{Y}}(z)~.
\end{equation} 
The simplest way to impose $W^3=0$ is to set to zero the total  RR term $\hat{F}=0$ \eq{RRtotal}, i.e.  not only the RR fluxes  $e_{\bf\L}=m^{\bf\L}=0$ but also the scalars $\z^A=\tilde\zeta_A=0$. 

From the ten-dimensional perspective, strictly speaking the pure spinor equations describing $\cN=1$ Minkowski and AdS vacua do not hold when $F=0$, as the resulting vacua are in fact $\cN=2$ \cite{gmpt3}. This argument uses the fact that only the RR term mixes the two spinors $\epsilon^1,\epsilon^2$ in the supersymmetry variations \rf{wv0}-\rf{int0}. Without this mixing there is no reason for one to not take the two four-dimensional components of the spinors to be independent parameters $\zeta_1$ and $\zeta_2$. This highlights the fact that the pure spinor equations derived above are formally conditions to have \emph{at least} $\cN=1$ supersymmetry, and in particular that $F\neq 0$ does not necessarily imply $\cN=1$ supersymmetry, as can be seen explicitly in certain orientifold examples \cite{gmpt3,Koerber:2007hd}. For domain wall backgrounds with $F=0$, this equates to the situation where $\cN=2$ supersymmetry is generated from two copies of $\cN=1$, meaning that the BPS projection conditions do not mix $\zeta_1$ and $\zeta_2$. It is then straightforward to check that, with an appropriately modified projector ansatz, the supersymmetry conditions can still be consistently written as in the previous section. In four dimensions, we are working in an $\cN=2$ language from the outset and the $F=0$ limit is perfectly consistent, being obtained by setting $\k=\l=0$.

Given the standard relation between the four- and ten-dimensional dilatons \eq{10Adil}, we need to evaluate $e^{-K_+}$ in order to determine the behaviour of $e^\phi$. One can proceed by rewriting the flow equation \eq{dXFSU33} in components 
\be
\partial_r\begin{pmatrix} {\rm Im}(h\,e^{U+\frac{K_+}2})\cr {\rm Im}(h\,e^{U+\frac{K_+}2}X^{i}) \cr {\rm Im }
(h\,e^{U+\frac{K_+}2}F_{0}) \cr {\rm Im }
(h\,e^{U+\frac{K_+}2}F_{i}) \end{pmatrix} =-\frac12e^{(1-p)U}\begin{pmatrix} G^{0}\cr G^i\cr G_{0}\cr G_i \end{pmatrix} ~,\label{bonobo}
\ee
where we have chosen $X^0=1$. As one can see from \eq{flux1}, $G_0$ contains the $H$ deformations\footnote{See appendix \ref{4dstr} for further explanation of the various deformations.}  $(e_0^A,\,e_{0A})$, i.e. the NS fluxes \eq{NSflux} and the $e_0$ component of the RR flux \eq{p5}. The torsion $T$  deformations $(e_i^A,\,e_{iA})$ \eq{decohSU3T} are all encoded in $G_i$, together with the $e_i$ components of the RR flux  \eq{p5}. $G^0$ contains the non-geometric deformations of type $R$ $(m_A^0,\,m^{0A})$ \eq{decohSU33} and the RR flux $m^0$ \eq{p5}. Finally, the $G^i$ contain the non-geometric deformations of type $Q$ $(m_A^i,\,m^{Ai})$ \eq{decohSU33} and the RR fluxes $m^i$ \eq{p5}. In the $SU(3)$ structure case $e_0$ is the IIA massive supergravity parameter, i.e. the Romans mass, $m^0$ corresponds to the Freund-Rubin parameter, $m^i$ are the two-form fluxes and $e_i$ are the four-form fluxes.

From the structure of \eq{bonobo} we can see that $e^{K_+}$ is extremely sensitive to the presence of $G^0$ and ${\rm Im}h$. In fact, for $G^0=0$ and ${\rm Im}h \neq 0$, we find that $e^{-K_+}=e^{2U}$. This case is particularly simple, as in the absence of RR terms (i.e. $W^3=0$) the ten-dimensional IIA dilaton $\phi$ is constant. In order to extract $e^{-K_+}$ one has to solve \eq{bonobo} case by case,  setting to zero the appropriate fluxes on the right-hand side. The most efficient way to achieve this is to first find the constraints imposed by the homogeneous equations, then to plug those with non-vanishing right-hand sides into \eq{Uprime2} and \eq{reW}, solving in terms of $U$. We shall not discuss all possible cases here, but rather focus on some interesting examples that make contact with the literature. We start by restricting ourselves to solutions with $W^0=W^i=0$, such that  \eq{bonobo} is considerably simplified. Such solutions occur in the absence of non-geometric deformations and for vanishing ``magnetic'' RR fluxes $m^{0}=m^i=0$.

The three cases of interest for us are:

\begin{itemize}

\item {\bf $T$ deformations} \\
For $G_i\neq0$ there is a unique solution with 
\be 
h=i,\quad\quad\quad e^{-K_+}=e^{2U},\quad\quad\quad b^i=0~.\quad\quad\label{T}
\ee 

\item {\bf $H$ deformations} \\
For $G_0\neq0$ there is a unique solution with 
\be 
h=1,\quad\quad\quad e^{-K_+}=e^{6U},\quad\quad\quad b^i={\rm const}~.\label{H}
\ee

\item {\bf $T+H$ deformations} \\
For $G_0\neq0,\,G_i\neq0 $ there is a unique solution with 
\be 
h=i,\quad\quad\quad e^{-K_+}=e^{2U},\quad\quad\quad b^i={\rm const}~.\label{HT}
\ee  

\end{itemize}
The fields $b^i$ are the real parts of the complex scalars $t^i=b^i + i v^i = \frac{X^i}{X^0}$. Let us stress that the above result is only sensitive to the presence of $G_0$ and $G_i$, and not to the fact that they contain $T$ deformations and $H$ fluxes, rather than ``electric'' RR fluxes. Consequently, \rf{T}-\rf{HT} hold when $F\neq 0$. However, this is not the case for the metric and dilaton, which are sensitive to the presence of the RR term $F$. Let us now rewrite the flow equations \eq{H1D}-\eq{H4D}  for $F=0$ and in the absence of non-geometric deformations. The metric always has the form \eq{metricF=0} while the structure of the flow equations depends on the presence of the torsion $T$ and the NS fluxes $H$.

\subsubsection*{$T$ deformations}

The pure $SU(3)$ structure case, that is when only $T$ deformations are present,  gives the standard Hitchin flow equations.
In the absence of $H$-fluxes and non-geometric $Q$ and $R$ fluxes, the generalised differential operator $\mathcal{D}$ reduces to the ordinary differential $d$. Using \eq{4dimdilF=0}-\eq{VGF=0} and \eq{T}, we find
\ba && \label{H1F=0}\partial_z {\rm Im}\Phi^0_-=-d{\rm Im}\Phi^0_+~,\\
&& \label{H2F=0}\partial_z {\rm Re}\Phi^0_+=-d{\rm Re}\Phi^0_- ~,\\
&& \label{H3F=0}d{\rm Im}\Phi^0_-=0~,\\
&&\label{H4F=0} d{\rm Re}\Phi^0_+=0~,
\ea
which are easily recognised to be the Hitchin flow equations describing a particular class of $SU(3)$ structure  six-manifolds that are known as half-flat \cite{Gurrieri:2002wz}. More explicitly, the necessary and sufficient conditions for an $SU(3)$ structure manifold to be half-flat are
\ba
&& d{\rm Re}\Phi^0_+\equiv -d (J\wedge J) = 0~, \\
&&d{\rm Im}\Phi^0_- \equiv d ({\rm Im} ~\Omega_\eta) = 0~.
\ea
The other two pure spinor equations give rise to
\ba
&& d{\rm Im}\Phi^0_+\equiv dJ = -\partial_z ({\rm Im}~\Omega_\eta) ,\\
&& d {\rm Re}\Phi^0_-\equiv d {\rm Re}~ \Omega_\eta =  -\frac12\partial_z(J\wedge J)~,
\ea
and imply that the total non-compact seven-manifold, constructed by the fibration of the six-manifold over the direction transverse to the domain wall, has $G_2$-holonomy. These equations were first  derived in the physics literature for a domain wall solution with $e_{0i}\neq0$ in \cite{Gurrieri:2002wz,Mayer:2004sd}. Here we have given their formulation with the general torsion compatible with the half-flat condition, and with constant ten-dimensional dilaton ($\phi=0$ for convenience).


\subsubsection*{$H$ deformations}
Let us now consider the case with non-vanishing NS H-flux in ten dimensions. The appropriate configuration which could give rise to a domain wall after compactification to four dimensions is generated by a stack of NS5-branes. As discussed in \cite{Gurrieri:2002wz}, it is possible to smear an NS5-brane over three of its four transverse directions such that the harmonic function depends on only one direction, which descends to the direction perpendicular to the domain wall in four dimensions. For such a configuration the ten-dimensional string frame metric, H-field and dilaton take the form
\begin{eqnarray}
ds^2  &=& ds^2_{\mathbb{R}^{1,2}}  + dz^2 + ds^2_Y(z) \label{ns5} ~,\\
\phi &=& \phi(z) ~,\nn \\
H&\in& H^3(\hat{Y}, \mathbb{R}) ~,
\end{eqnarray}
where the flux $H$ is harmonic \eq{notadSU3}. When $Y$ is Calabi-Yau, the mirror manifold $\tilde Y$ is precisely the half-flat, $SU(3)$ structure manifold with vanishing H-flux and constant dilaton \cite{Gurrieri:2002wz} discussed in the previous subsection. In this case $Y$ has SU(3) holonomy and the pure spinors $\Phi_\pm$ are the familiar K\"ahler form $J$ and holomorphic three-form $\Omega$, both of which are closed. The transverse flow of the pure spinors is then supported solely by the H deformations. The metric \eq{ns5} evidently coincides with \eq{metricF=0}, while $H$ is harmonic by construction \eq{NSflux}. From \eq{H} and \eq{10Adil}, we find that the ten-dimensional dilaton is $e^\phi=e^{2U}$. In this case the domain wall flow equations become
\begin{eqnarray}
d_H \mathrm{Im}\left[e^{-\phi}~ \Phi^0_- \right] &=& 0 \label{ps1f0} ~,\\
d_H \mathrm{Re}\left[e^{-\phi}~\Phi^0_- \right] &=& - \pt_z \mathrm{Im} \left[ e^{-\phi}~\Phi^0_+ \right] \label{ps2f0} ~,\\
d_H \mathrm{Im}\left[e^{-\phi}~  \Phi^0_+ \right] &=& 0  \label{ps3f0}~,\\
d_H \mathrm{Re}\left[e^{-\phi}~\Phi^0_+  \right] &=& \pt_z \mathrm{Im} \left[ e^{-\phi}~  \Phi^0_- \right] \label{ps4f0}~.
\end{eqnarray}

\subsubsection*{$T+H$ deformations}
We shall now consider the most generic geometric background with $F=0$. As one can see from equations \eq{T} and \eq{HT}, it is quite similar to the $T$ background, apart from the fact that the NSNS two-form scalars can take any constant value, not necessarily zero. The ten-dimensional dilaton is constant and we choose $\phi=0$ for simplicity. The flow equations are then 
\ba && \label{TH1F=0}\partial_z {\rm Im}\Phi^0_-=-d_H{\rm Im}\Phi^0_+~,\\
&& \label{TH2F=0}\partial_z {\rm Re}\Phi^0_+=-d_H{\rm Re}\Phi^0_- ~,\\
&& \label{TH3F=0}d_H{\rm Im}\Phi^0_-=0~,\\
&&\label{TH4F=0} d_H{\rm Re}\Phi^0_+=0~.
\ea
This set of equation were first derived in \cite{Witt}, where it was shown that they describe the embedding of an $SU(3)\times SU(3)$ structure manifold into a $G_2\times G_2$ structure manifold. In particular, it was shown that the flow equations lift to a set of conditions which an describe a generalised $G_2$ structure $\rho~,\hat\rho$ on $M_7=\hat{Y}\times \mathbb{I}_z$ which is integrable with respect to $H$, as we described in detail above (see \rf{r} and \rf{rh}). A particular case of such backgrounds with only $e_{0\bf\L}\neq0$ is included in the analysis of \cite{Louis:2006wq}. In the absence of the flow terms (i.e. the $\pt_z$ terms), and if $H$ is a primitive $(2,1)$-form, then these equations are equivalent to Gualtieri's definition of a twisted generalised K\"ahler structure (see section 6 of \cite{Gualtieri:2003dx}).


\subsection{Non-geometric deformations}

Finally, we shall comment briefly on the case of non-geometric backgrounds. From the four-dimensional perspective, whenever $W^3=0$ the dilaton $\varphi$ and the metric take the form \eq{4dimdilF=0} and \eq{metricF=0},  even in the presence of non-geometric deformations. As a consequence of \eq{VGF=0}, the flow equations \eq{H1D}-\eq{H4D} can then be written as
\begin{eqnarray}
\mathcal{D}\mathrm{Im}\left[e^{-\phi}~ \hat\Phi^0_- \right] &=& 0 \label{ps1f0n} ~,\\
\pt_z \mathrm{Im} \left[ e^{-\phi}~h~\hat\Phi^0_+ \right] &=&-\mathcal{D}\mathrm{Re}\left[e^{-\phi}~\hat\Phi^0_- \right] \label{ps2f0n} ~,\\
\pt_z \mathrm{Im} \left[ e^{-\phi}~  \hat\Phi^0_- \right] &=&\mathcal{D} \mathrm{Re}\left[e^{-\phi}~h~\hat\Phi^0_+  \right] ~, \label{ps3f0n}\\
\mathcal{D} \mathrm{Im}\left[e^{-\phi}~h ~ \hat\Phi^0_+ \right] &=& 0  \label{ps4f0n}~,
\end{eqnarray}
where the precise expression for the dilaton $\phi$ can be obtained by using \eq{bonobo} to determine $K_+$, analogously to the geometric case. We shall limit ourselves to the example of $Q$ deformations, for which we find a solution with 
\begin{equation} 
h=1,\quad\quad\quad e^{-K_+}=e^{-2U},\quad\quad\quad b^i=0~, \label{Q}
\end{equation}
and, therefore, the dilaton in \rf{ps1f0n}-\rf{ps4f0n} is given by $e^\phi=e^{-2U}$.

Turning to the comparison of the flow equations, we recall that in ten dimensions our derivation assumed that the background was globally geometric. Nevertheless, it has been argued \cite{Cassani:2007pq} that one can formally incorporate non-geometric charges in the pure spinor equations for $\cN=1$, maximally symmetric vacua by replacing the twisted derivative $d_H$ appearing there with the generalised derivative $\mathcal{D}$ \rf{defD}. We shall not pursue the non-geometric case in any detail here, but note that on substituting $d_H \rightarrow \mathcal{D}$ in the domain wall pure spinor equations derived in ten dimensions \rf{nps1}-\rf{nps4}, we find formal agreement with the four-dimensional result presented above \rf{ps1f0n}-\rf{ps4f0n}.


\section{Discussion}\label{disc}

We have studied BPS domain wall configurations in gauged four-dimensional supergravity arising from type II supergravity compactified on an $SU(3) \times SU(3)$ structure manifold. Starting in four dimensions, we used standard manipulations of the supersymmetry transformations to derive a set of flow equations for the scalar fields of the vector and hypermultiplets in a domain wall background. We then showed how these equations could be recast as a set of generalised Hitchin flow equations, describing the embedding of the $SU(3) \times SU(3)$ structure manifold into a $G_2 \times G_2$ structure, or generalised $G_2$, manifold, provided that the pure spinors satisfied $\langle \mathcal{D} {\rm Im}\hat\Phi^0_-,\,\hat\Phi^0_+\rangle=0$. Interestingly, from the ten-dimensional perspective, this condition follows directly from one of the supersymmetry constraints for a domain wall configuration \rf{nps} and the compatibility constraint \rf{comp} for an $SU(3) \times SU(3)$ structure manifold.

For simplicity, our ten-dimensional analysis focused solely on configurations that could give rise to domain walls in four dimensions preserving at least two supersymmetries. This allowed us to adapt the formalism previously used to describe maximally symmetric type II supergravity vacua in terms of pure spinors. As we have already noted, the conditions of Gra\~na et al. \cite{gmpt2} are strictly for backgrounds preserving \emph{at least} $\cN =1$ supersymmetry in four dimensions. The same applies to the domain wall configurations here, and by carefully comparing our ten-dimensional result with the orientifold truncation of the four-dimensional counterpart, we were able to show a precise agreement between the two approaches.  

This matching between the equations describing domain wall vacua in the ten-dimensional and four-dimensional theories is a useful test of the $SU(3) \times SU(3)$ structure compactification procedure proposed in \cite{Grana:2006hr}. For maximally symmetric vacua this check was carried out in \cite{Cassani:2007pq}. As we argued above, due to the prevalence of domain wall vacua in gauged supergravities our results provide a valuable additional check. Furthermore, the generalised Hitchin flow equation for domain walls derived here are symmetric under the proposed generalisation of mirror symmetry for $SU(3) \times SU(3)$ structure backgrounds:
\be
\hat \Phi^0_+ \leftrightarrow \hat\Phi^0_-~,\quad\quad F_{IIA} \leftrightarrow F_{IIB}~.
\ee

The flow equations we derived in ten dimensions also included a non-trivial warp factor $A$. However, in order to make a strict comparison with our four-dimensional result we made the standard assumption that the warp factor vanishes \cite{Grana:2006hr}. It would be interesting to reconsider domain wall vacua in warped compactifications directly in terms of $\cN=1$ supergravity in four dimensions. In \cite{ten2four} it was suggested that the appropriate four-dimensional effective theory for warped $SU(3) \times SU(3)$ compactifications is a partially gauge fixed version of matter-coupled, $\cN=1$ superconformal supergravity (see also \cite{hlmt}). However, the relation between this and the warped version of the off-shell approach of \cite{Grana:2006hr} remains unsettled. In future work, we aim to reassess `warped' domain wall vacua in the $\cN=1$ superconformal theory and determine their relation to the cases we have considered here \cite{SV2}. 

Finally, we shall briefly comment on the applications of the generalised Hitchin flow equations we have derived to gauge/gravity duality. Recently, it has been realised that Chern-Simons-matter conformal field theories are dual to massive type IIA supergravity solutions of the form $AdS_4 \times \mathbb{CP}^3$, with the Romans mass $F_0$ acting as a deformation parameter (see \cite{Gaiotto:2009yz,Petrini:2009ur} and references therein). It has been suggested in \cite{Gaiotto:2009yz} that the appropriate equations for describing the gravity duals of such Chern-Simons-matter theories should be a generalised version of the Hitchin flow equations. Therefore, our results \rf{ps1}-\rf{ps4} should prove useful in constructing interesting examples of these gravity duals.

\begin{center}
{\large  {\em Acknowledgements}}
\end{center}

We would like to thank P. Koerber, J. Louis, P. Meessen, M. Petrini, V. Stojevic and A. Tomasiello for discussions, and L. Martucci for his collaboration in the early stages of this work. We especially thank  R. Reid-Edwards for useful discussions and his careful reading of this manuscript. S.V. is partially supported by the Spanish MEC grant FPA2006-00783, a MEC Juan de la Cierva scholarship, the CAM grant HEPHACOS P-ESP-00346 and the Spanish Consolider-Ingenio 2010 program CPAN CSD2007-00042. P.S. is supported by the German Science Foundation (DFG) and would also like to thank the K. U. Leuven for its support at various stages. 

This article is dedicated to the memory of Raffaele Punzi, a friend and colleague who will be missed.

\vskip 2cm

\appendix

\section{Conventions}\label{conv}

The gamma matrices in ten dimensions satisfy $\{\g_M,\g_N\} = 2g_{MN}$. In four dimensions we use the mostly-minus convention for the metric, making it easier to compare with the gauged supergravity literature, whereas in ten dimensions it is more convenient to use the mostly-plus convention. The details of how to change between conventions can be found in section \ref{s4}.  

The covariant derivative on spinors is defined as $D_M = \pt_M + \frac{1}{4}\O_M^{~\ul{AB}}\G_{\ul{AB}}$, where flat indices appear underlined. All four- and six-dimensional gamma matrices are Hermitian, apart from $\gamma^0 = -\gamma^\dagger$. The four-dimensional gamma matrices are chosen to be pure real, while the six-dimensional gamma matrices are pure imaginary. With these definitions, the chirality matrices are 
\begin{equation}
\begin{array}{ccccccc}
\gamma_5 & = & i\gamma_{\ul{0}} \ldots \gamma_{\ul{3}} & = & - \gamma_5^* & = &
\gamma_5^{\dagger} ~,\\
\gamma_7 & = & -i\gamma_{\ul{4}} \ldots \gamma_{\ul{9}} & = & - \gamma_7^* & = &
\gamma_7^{\dagger} ~.
\end{array}
\end{equation}
Note that with our conventions $\gamma_5$ is pure imaginary.
The ten-dimensional gamma matrices can be decomposed as
\begin{equation}
\Gamma_{\mu} = \gamma_{\mu} \otimes 1 , \qquad
\Gamma_m=\gamma_5\otimes\gamma_m~,
\end{equation}
where $\mu$ labels four-dimensional directions and $m$ labels six-dimensional directions on the internal manifold. The ten-dimensional chirality matrices are
\begin{equation}
\Gamma_5=\gamma_5\otimes 1~, \qquad \Gamma_7 = 1\otimes\gamma_7~, \qquad
\Gamma_{11} = \gamma_5 \otimes \gamma_7 = \Gamma_{11}^{\dagger} .
\end{equation}

We often use the democratic formalism \cite{bkorp}, where all spinor indices suppressed. For instance, $\epsilon = (\epsilon^1, \epsilon^2)$ is a doublet of ten-dimensional Majorana-Weyl (MW) spinors. For type IIA backgrounds the supersymmetry parameter is a ten-dimensional Majorana spinor $\epsilon$ that can be split in two MW spinors of opposite chirality: 
\bea
\epsilon=\epsilon_1 +\epsilon_2~,\quad\quad
\Gamma_{(10)}\epsilon_1=\epsilon_1 ~,\quad\quad
\Gamma_{(10)}\epsilon_2=-\epsilon_2\ . 
\eea 
In type IIB the two supersymmetry parameters $\epsilon_{1,2}$ are MW spinors of positive ten-dimensional chirality ($\Gamma_{(10)}\epsilon_{1,2}=\epsilon_{1,2}$). 

\section{Pure spinor definitions}\label{psd}

Recall that generalised complex geometry studies structures on the direct sum of the tangent and cotangent bundles $T_{\hat{Y}}\oplus T^\star_{\hat{Y}}$ \cite{gmpt1,gmpt2,Hitchin:2004ut,Gualtieri:2003dx,Witt}. A generalised almost complex structure $\cal J$ is defined as a map of $T_{\hat{Y}}\oplus T^\star_{\hat{Y}}$ onto itself such that ${\cal J}^2 = -\mathbb{I}$, and obeys the Hermiticity condition ${\cal J}^t \d \cal J = \d$, where $\d = \binom{0 \ \ 1} {1 \ \ 0}$ is the natural metric on $T_{\hat{Y}}\oplus T^\star_{\hat{Y}}$. The existence of $\d$ implies that the structure group on $T_{\hat{Y}}\oplus T^\star_{\hat{Y}}$ is reduced to $O(6,6)$. Spinors lie in representations of the $\Spin(6,\!6)$ cover group, although often one refers to the related representation of the Clifford algebra $\mathrm{Clifford(6,\!6)}$, which can be defined in terms of matrices $\l^m, \r_n$ obeying the following algebra \cite{gmpt1,gmpt2}
\be
\{\l^m,\l^n \}=0~,\quad\quad\{\l^m,\r_n\}=\d^m_n~,\quad\quad\{\r_m,\r_n\}=0~,
\ee
where $\d^m_n$ is the $6+6$-dimensional metric on $T_{\hat{Y}}\oplus T^\star_{\hat{Y}}$, described above, and $m,n=1,\ldots,6$. One can also find a representation of this algebra in terms of forms using
\be
\l^m = dx^m\w~,\quad\quad\r_n = \iota_n~,
\ee
where $\iota_n\equiv \iota_{\pt_n} dy^{a_1}\w\ldots \w dy^{a_p} = p\d_n^{[a_1}dy^{a_2}\w\ldots \w dy^{a_p]}$ is the familiar contraction $\iota_n : \L^pT^\star \rightarrow \L^{p-1}T^\star$. The spaces of positive and negative chirality $\Spin(6,\!6)$ spinors are seen to be isomorphic to spaces of even and odd degree differential forms on ${\hat{Y}}$. A pure spinor $\chi$ is annihilated by a six-dimensional subspace of $\mathrm{Clifford(6,\!6)}$ and is equivalent to a {\it polyform}, or formal sum of forms, of even or odd degree. Furthermore, one can show that there is a one-to-one correspondence between the line bundle of pure spinors and generalised almost complex structures. We refer the reader to \cite{gmpt3} for further details and to \cite{Grana:2005ny} for details of the isomorphism between $\mathrm{Spin(6,\!6)}$ spinors and forms.

We shall make much use of the Clifford map, which relates $\mathrm{Spin(6,\!6)}$ spinors to $\mathrm{Spin(6)}$ bispinors \cite{gmpt1,Witt}:
\bea\label{cmap}
\chi\equiv\sum_k \frac{1}{k!}\chi^{(k)}_{m_1\ldots m_k} dx^{m_1}\w\ldots\w dx^{m_k}\longleftrightarrow\ \slashchar{\chi}\equiv\sum_k \frac{1}{k!}\chi^{(k)}_{m_1\ldots m_k}\gamma^{m_1\ldots m_k}~,
\eea 
where $\g^m$ are the $\mathrm{Clifford(6)}$ gamma matrices defined in appendix \ref{conv} and $g_{mn}$ is the metric on ${\hat{Y}}$. Purity then reduces to saying that 3 linear combinations of the gamma matrices annihilate $\slashchar{\chi}$. 

We will also use the Mukai pairing between forms, which is antisymmetric in six dimensions. Using the Clifford map, the Mukai pairing can be defined in terms of the spinor trace over the product of bispinors $\slash A$ and $\slash B$
\begin{equation}
\langle A_k, B_{6-k} \rangle = \frac{(-1)^k}{8}\mathrm{Tr}( \slas{\star A_k} \slash{B_{6-k}})\mathrm{vol_6}~,\label{Mukai}
\end{equation}
where $\mathrm{vol_6}$ is the volume of ${\hat{Y}}$. For pure spinors, which are polyforms, the Mukai pairing selects only the top degree forms i.e. it is always proportional to $\mathrm{vol_6}$: 
\begin{equation}
\langle A, B \rangle~\mathrm{vol}_6 \equiv \left( A \wedge \sigma(B) \right)|_\mathrm{top}~,
\end{equation}
where we use the involution $\sigma$ which reverses the order indices on a form. For a collection of useful Mukai pairing properties, see appendix B of \cite{Cassani:2007pq}.

For the backgrounds we are interested in we can decompose the ten-dimensional spinors as in \rf{spinors} (for type IIA) and associate two pure spinors to the internal spinor components $\eta_+^{(1)}$  and $\eta_+^{(2)}$, which for now we will take to have unit norm: 
\bea
\slashchar{\chi}_+=\eta^{(1)}_+\otimes\eta^{(2)\dagger}_{+}~,\quad\quad \slashchar{\chi}_-=\eta^{(1)}_+\otimes\eta^{(2)\dagger}_{-}~. \label{unps}
\eea
We can rewrite these bispinors in terms of polyforms of definite parity using the Fierz identities:
\be
\eta^{(1)}_+\otimes\eta^{(2)~\dagger}_\pm = \frac{1}{4} \sum_k \frac{1}{k!} \eta^{(2)~\dagger}_\pm \gamma^{m_1\ldots m_k}  \eta^{(1)}_+ \gamma^{m_k\ldots m_1}~.
\ee
It is important to note that the slash and conjugation do not commute for pure spinors:
\bea
\overline{(\slashchar{\chi}_+)}=\slashchar{\overline{\chi}}_+=\eta^{(1)}_-\otimes\eta^{(2)\dagger}_-~,\quad\quad \overline{(\slashchar{\chi}_-)}= -\slash{\overline{\chi}}_-  =\eta^{(1)}_-\otimes\eta^{(2)\dagger}_+~.
\eea
When the context is clear, we shall drop the slash on pure spinors for notational convenience. 

Given a pair of pure spinors $\chi_\pm$, they will each define an $SU(3,3)$ structure on $T_{\hat{Y}}\oplus T^\star_{\hat{Y}}$. If, in addition, they obey the following set of compatibility and normalisation conditions they define an $SU(3)\times SU(3)$ structure. First note that an element $V=v + \xi$ of  $T_{\hat{Y}}\oplus T^\star_{\hat{Y}}$ acts on a polyform as
\be
V \cdot \chi = \iota_v \chi + \xi \wedge \chi~.
\ee
Compatibility is then given in terms of the Mukai pairing as
\begin{equation}
\langle \chi_{+}, V\cdot \overline{\chi}_{-} \rangle =  0 = \langle \chi_{+}, V\cdot {\chi}_{-} \rangle~, \label{comp}
\end{equation}
for any $V\in T_{\hat{Y}}\oplus T^\star_{\hat{Y}}$, and pure spinors constructed as in \rf{unps} are automatically compatible. The normalisation condition is 
\begin{equation}\label{spinnorm}
\langle \chi_{+}, \overline{\chi}_{+} \rangle = \langle \chi_{-}, \overline{\chi}_{-} \rangle~.
\end{equation}

In \cite{gmpt2} it was shown how the type II supersymmetry transformations can be written in an elegant form in terms of pure spinors using the Clifford map. The same manipulations can be applied to the domain wall backgrounds we are interested and we refer to \cite{gmpt2,gmpt3} for further details of these calculations. The obvious difference, however, is the appearance of a transverse r-dependence in the domain wall metric \rf{metric}. Following \cite{hmt}, we shall choose a local frame on ${\hat{Y}}$ such that the vielbein $e_{\ul{a}}^m$ obey $(e_{\ul{a}}{}^m)' =  -\frac12 g^{mp} g_{pn}' e_{\ul{a}}{}^n$, where  $g_{pn}'$ is symmetric in its indices. In order to manipulate the terms involving transverse derivatives we then use the following identity
\begin{equation}
\partial_r(\slashchar{\chi}_{_\pm})= \!\!\!\!
\begin{picture}(10,10)(-15,5)
\put(0,0){\line(6,1){100}}
\end{picture}
(\partial_r \chi_{_\pm} - \half g^{mp}g'_{pn} dx^n\wedge \iota_m \chi_{_\pm})
=\!\!\!\! \begin{picture}(10,10)(-15,0)
\put(0,0){\line(2,1){20}}
\end{picture}
(\partial_r \chi_{_\pm}) - \frac{g'_{mn}}{4} \left( g^{mn}\slashchar{\chi}_{_\pm} \mp \frac12 
\gamma^m \slashchar{\chi}_{_\pm}\gamma^n\right)\, .
\end{equation}


\newpage

\section{$\cN=2$ supersymmetry transformation laws}\label{fermtrans}

Here we shall explicitly state the supersymmetry transformation laws for the $\cN=2$ theory described in section \ref{s2}, taking advantage of the results \cite{Andrianopoli:1996cm,Dall'Agata:2003yr,Sommovigo:2004vj,D'Auria:2004tr,D'Auria:2007ay}. They take the following form:
\ba
\d\p_{\m {\hat{A}}}&=&D_\m\ve_{\hat{A}}-\frac i2\tilde{M}^{\bf \L\S}\tilde{H}_{{\bf\L}\m}\vec{\o}_{\bf \S}\cdot\vec\s_{{\hat{A}}}^{\ {\hat{B}}}\ve_{\hat{B}}+\left[iS_{{\hat{A}}{\hat{B}}}\eta_{\m\n}+\e_{{\hat{A}}{\hat{B}}}T^-_{\m\n}\right]\g^\n\ve^{\hat{B}}~,\nn\\
\d\l^{i{\hat{A}}}&=&i\partial_\m t^i\g^\m\ve^{\hat{A}}+\e^{{\hat{A}}{\hat{B}}}G^{i\,-}_{\m\n}\g^{\m\n}\ve_{\hat{B}}+W^{i{\hat{A}}{\hat{B}}}\ve_{\hat{B}}~,\nn\\
\d\z_{\hat\a}&=&iP_{u {\hat{A}}\hat\a}\partial_\m q^u\g^\m\ve^{\hat{A}}-i\tilde{M}^{\bf \L\S}\tilde{H}_{{\bf\L}\m}\mathcal{U}_{{\bf \S}\,{\hat{A}}\hat\a}\g^\m\ve^{\hat{A}}+N_{\hat\a}^{\hat{A}}\ve_{\hat{A}}~,\nn\\
&&\label{ferm}\\
\d V^a_\m&=&-i\pb_{{\hat{A}}\m}\g^a\ve^{\hat{A}}-i\pb^{\hat{A}}_\m\g^a\ve_{\hat{A}}~,\nn\\
\d A^{\bf\L}_\m &=& 2 L^{\bf\L} \overline{\psi}^{\hat{A}}_\m \ve^{\hat{B}} \e_{{\hat{A}}{\hat{B}}} + 2 \overline{L}^{\bf\L} \overline{ \psi}_{{\hat{A}}\m}
\ve_{\hat{B}} \e^{{\hat{A}}{\hat{B}}} \nn\\
&&+\left( i f^{\bf\L}_k \overline{ \l}^{k{\hat{A}}} \g_\m \ve^{\hat{B}} \e_{{\hat{A}}{\hat{B}}} + i \overline{f}^{\bf\L}_{\bar k} \overline{\l}^{\bar
k}_{\hat{A}} \g_\m \ve_{\hat{B}} \e^{{\hat{A}}{\hat{B}}}\right)~,\nn\\
\d B_{{\bf\L}\m\n} &=& - \frac i2 \left(\overline{\ve}_{\hat{A}} \g_{\m\n} \z_{\hat\a} \op{U}_{\bf \L}{}^{{\hat{A}}\hat\a} -
\overline{ \ve}^{\hat{A}} \g_{\m\n} \z^{\hat\a} \op{U}_{{\bf\L}{\hat{A}}\hat\alpha}\right)\nn\\
&& -\frac i2 \vec \o_{{\bf\L}}\cdot\vec\s_C{}^{{\hat{A}}} \left( \overline{\ve}_{\hat{A}} \g_{[\m}\psi^C_{\n]}+ \overline{\psi}_{[\m {\hat{A}}}
\g_{\n]} \ve^C\right) ~,\nn\\
\d t^k &=& \overline{\l}^{k{\hat{A}}}\ve_{\hat{A}} ~,\nn\\
\d q^u &=& P^u{}_{{\hat{A}}\hat\a} \left(\overline{\z}^{\hat\a} \ve^{\hat{A}} + \mathbb{C}^{\hat\a\hat\b}\e^{{\hat{A}}{\hat{B}}}\overline{\z}_{\hat\b} \ve_{\hat{B}}
\right)~.\nn\\
&&\label{bos}
\ea

In the case of an $SU(3)$ structure compactifications there are no massive tensor multiplets, that is \eq{ferm} and \eq{bos} describe a standard $\cN=2$ gauged supergravity \cite{Andrianopoli:1996cm} . This is tantamount to saying that all the tensor multiplets can be dualised into scalars. More specifically, it implies $\tilde{H}_{{\bf\L}\m}=B_{{\bf\L}\m\n}=0$ and that  $P^u{}_{{\hat{A}}\hat\a}$ is the vielbein of the quaternionic manifold parameterised by the hyperscalars $q^u$. The specific gauging is Abelian and it is described in \cite{D'Auria:2004tr}. The fermion shifts $S_{{\hat{A}}{\hat{B}}},\,W^{i\,{\hat{A}}{\hat{B}}},\,N^{\hat{A}}_{\hat\a}$  are completely determined by specifying the triplet of superpotentials $\vec W$ \eq{SU2W}-\eq{N}. Their explicit form is given in \eq{WSU3}.

In the case of an $SU(3)\times SU(3)$ structure compactification \cite{D'Auria:2007ay} the theory features $h^{(1,1)}+1$ tensors 
\be B_{{\bf\L}\m\n}\equiv e^I_{\bf\L}B_{I\,\m\n}~,\ee
where  $e^I_{\bf\L}\equiv(e^A_{\bf\L},\,e_{A\bf\L})$ and $B_{I\,\m\n}\equiv(B^A_{\,\m\n},\,B_{A\,\m\n})$ are dual to the RR scalars. It is important to note that the combination of RR scalars appearing in the fermions shifts \eq{WSU3} is gauge invariant \cite{D'Auria:2007ay}, that means that it is not the combination of scalars that is dualised into tensors. For this reason the flow equation \eq{attq0} has the same structure in both  $SU(3)$ and $SU(3)\times SU(3)$ structure case.

The tensors appear in the fermionic supersymmetry transformation laws \eq{ferm} through the Hodge dual of their field strength $\tilde{H}_{\bf\L}=\star {dB}_{\bf\L}$. The scalars $q^u$  that have not  been dualised are parameterised by the rectangular vielbein $P^u{}_{A\a}$   \cite{Dall'Agata:2003yr}. We have defined for convenience
\be \tilde{M}^{\bf \L\S}\equiv\tilde{e}_I^{\bf\L}\tilde{e}_J^{\bf\S}M^{IJ}~, \ee
where $ \tilde{e}_I^{\bf\L}e^I_{\bf\S}=\d_{\bf\L}^{\bf\S}$. Similarly we define
\ba
\vec\o_{\bf\L}&\equiv &e^I_{\bf\L}\vec\o_I~,\\
\mathcal{U}^{A\hat\a}_{\bf\L}&\equiv&e^I_{\bf\L} \mathcal{U}^{A\hat\a}_I~.
 \ea
The triplet of superpotentials $\vec W$ can be derived from \cite{D'Auria:2007ay}   using the techniques described in \cite{Dall'Agata:2003yr} and is given by \eq{WSU33}.


\section{4D $\cN=2$ supergravity and $SU(3)\times SU(3)$ structures}\label{4dstr}

In this appendix we will briefly review $SU(3)$ and $SU(3)\times SU(3)$ compactifications from the perspective of four-dimensional $\cN=2$ supergravity. For further details and a complete discussion of the Kaluza-Klein procedure for $SU(3)$ and $SU(3)\times SU(3)$ structure manifolds, we refer the reader to \cite{Grana:2005ny,Grana:2006hr}.

\subsection*{$SU(3)$ structure}

When $Y$ has $SU(3)$ structure  or $SU(3)$ holonomy, the pure spinors are defined in terms of the complex three-form 
\be\O=Z^A\a_A-G_A\b^A\label{Omega}~,\ee
and the real K\"ahler two-form $J$, which is often combined with the $B$-field into a complex two-form
\be B+iJ=\frac{X^i}{X^0}\,\o_i~,\ee
where
\be (\o_i\,,\,\o^i)\quad i=1,\dots h^{(1,1)},\quad\quad (\a_A\,,\,\b^A)\quad A=0,1,\dots h^{(2,1)}~, \label{asu3basis}\ee
and $(X^0, ~X^i)$ and $Z^A$ are homogeneous complex coordinates for the K\"ahler and complex structure moduli, respectively. In the Calabi-Yau case, \rf{asu3basis} are the usual basis of harmonic forms. 

We are not going to consider the most generic  $SU(3)$ structure and $SU(3)\times SU(3)$ manifolds but we will restrict to those that, according to \cite{Grana:2005ny}, reduce to the $\cN=2$ ungauged theory in the limit of vanishing gauge coupling.  This is achieved by deleting the triplet $\mathbf 3$ and $\mathbf{\overline{3}}$ representations in the  $SU(3)$ and $SU(3)\times SU(3)$  field decomposition  \cite{Grana:2005ny,Grana:2006hr}. For the $SU(3)$ structure case this means that the torsion classes $\mathcal{W}_4$ and $\mathcal{W}_5$ must vanish \cite{CS}. The failure of the basis of forms to be closed can then be expressed as 
\ba 
d\a_A=e_{A{i}}\,\o^{i},\quad\quad  d\b^A=e^A_{i} \o^{i}~,\nn\\
d\o_{i}=e_{i}^A\a_A-e_{A{i}}\b^{A},\quad\quad d\o^{i}=0\label{decohSU3T}~,
\ea
where we do not write the terms that vanish under the Mukai pairing.  

It is convenient to introduce a zero-form $\o_0\equiv1$ and a six-form $\o^0$ such that $\int\o^0=-1$ and an index ${\bf\L}=(0,\,i)$, ${\bf\L}=0,1,\dots h^{(1,1)}$ in order to allow us to write the two pure spinors as
\ba
&&\Phi^0_+=X^{\bf\L}\o_{\bf\L}-F_{\bf\L}\o^{\bf\L}\label{aPhi0+}~,\\
&&\Phi^0_-=Z^A_\eta\a_A-G_{\eta\,A}\b^A\label{aPhi0-}~,
\ea
that is
\bea
\Phi^0_+= X^0 e^{B+iJ}~,\quad\quad \Phi^0_-=\eta\Omega\ , \label{Phisu3}
\eea
where $X^0$ can be fixed to any constant phase, while $\eta$ is a normalisation factor that we shall fix later \eq{eta}. 

In \eq{aPhi0+} $(X^{\bf\L},\,F_{\bf\L})$ is a symplectic section and $F_{\bf\L}=\frac{\partial F}{\partial X^{\bf\L}}$, where $F$ is the holomorphic prepotential for the K\"ahler moduli defined by
\be 
F=-\frac1{3!}\,\mathcal{K}_{ijk}\frac{X^iX^jX^k}{X^0}~.
\ee 
The scalars $t^i=\frac{X^i}{X^0}$ parameterise the special K\"ahler manifold of the vector multiplet sector of the resulting four-dimensional theory, with the K\"ahler potential given by
\be 
\label{JJJ}e^{-K_+}=\frac43\int J\wedge J\wedge J~.
\ee
Similarly  in \eq{Omega}  $(Z^A,\,G_A)$ is a symplectic section and $G_A=\frac{\partial G}{\partial Z^A}~,$ where $G$ is the holomorphic prepotential for the complex structure moduli defined by
\be 
G=-\frac1{3!}\,d_{abc}\frac{Z^a Z^b Z^c}{Z^0}~.
\ee 
The  scalars $z^a=\frac{Z^a}{Z^0}$ parameterise a special K\"ahler submanifold of the quaternionic manifold in the hypermultiplet sector of the four-dimensional theory, with K\"ahler potential $K_-$ defined as
\be 
\label{OO}e^{-K_-}=i\int\O\wedge\bar\O~.
\ee
The rescaled sections introduced in \eq{aPhi0-} are defined as
\be
(Z^A_\eta\,,G_{\eta A})\equiv \eta (Z^A\,,G_A)\label{aZresc}~,
\ee
where the normalising factor is given by
\be
\label{eta}\eta=e^{\frac{K_--K_+}2}~.
\ee
The rescaling \eq{Zresc}  is performed in order to give the correct normalisation of the pure spinors \eq{spinnorm}. As we shall describe later, it is often more convenient to define twisted differential, rather than use the twisted differential form described above. For vanishing $B$ field, $J$ and $\Omega$ satisfy the usual $SU(3)$ structure conditions
\bea
J\wedge \Omega=0 \, , \qquad \frac{1}{3!}J\wedge J\wedge J=-\frac{i}{8}\eta^2\Omega\wedge\bar\Omega.\label{JO}
\eea
The conditions \eq{JO} are the equivalents of \rf{comp} and \rf{spinnorm}.

\subsubsection*{$T$ and $H$ deformations}

We will refer in the following to the $(e^A_i,\,e_{Ai})$ deformations as $T$ deformations, since they are all geometric deformations induced by the torsion. As we want to consider the presence of NS fluxes
\be
\label{aNSflux}H=e_0^A{\a}_A-e_{0A}\b^A~,
\ee
it is convenient to define a twisted derivative operator $ d_H\equiv d-H\wedge$,
such that on the $SU(3)$ structure manifold basis, extended with $(\o_0,\,\o^0)$, we have
\ba 
d_H\a_A=e_{A{\bf\L}}\o^{\bf\L},\quad\quad  d_H\b^A=e^A_{\bf\L} \o^{\bf\L}~,\nn\\
d_H\o_{\bf\L}=e_{\bf\L}^A\a_A-e_{A{\bf\L}}\b^{A},\quad\quad d_H\o^{\bf\L}=0\label{adecohSU3TH}~.
\ea
We will refer to $(e_0^A,\,e_{0A})$ as $H$ deformations. Note that $d^2=d_H^2=0$ implies
\be 
e_{A{\bf\L}}e^A_{~\bf\S}-e_{A{\bf\S}}e^A_{~\bf\L}=0\label{anotadSU3}~.
\ee


\subsection*{$SU(3) \times SU(3)$ structure} 

When $\hat{Y}_6$ is an $SU(3)\times SU(3)$ structure manifold, both the even and odd basis forms on the truncated spaces of forms
\be (\hat\o_{\bf\L},\,\hat\o^{\bf\L})\quad {\bf\L}=0,1,\dots h^{(1,1)}\quad , \quad  (\hat\a_A\,,\,\hat\b^A)\quad A=0,1,\dots h^{(2,1)}~,\label{aPFbasis}\ee
are defined as polyforms, such that both pure spinors are polyforms and are given by
\ba
&&\hat\Phi_+^0=X^{\bf\L}\hat\o_{\bf\L}-F_{\bf\L}\,\hat\o^{\bf\L}\label{ap1}~,\\
&&\hat\Phi_-^0=Z^A_\eta\hat\a_A-G_{\eta A}\,\hat\b^A\ .\label{ap2}
\ea
The K\"ahler potentials are defined by
\be
e^{-K_\pm} = i \int \langle \hat\Phi_\pm^0~,~ \overline{\hat\Phi_\pm^0} \rangle~. \label{genkp}
\ee
Hatted quantities are introduced to distinguish the more general $SU(3)\times SU(3)$ structure case. In order to include the RR fields, we define a third polyform $\hat\S$ 
\be
\hat\S=\zeta^A\hat\alpha_A-\tilde\zeta_{ A}\hat\beta^A~,\label{ap3}
\ee
where $(\z^A,\tilde\z_A)$ are the RR scalars, that in the four-dimensional theory belong to the hypermultiplet sector. When the theory is truncated to an $\cN=1$ spectrum it is further convenient to introduce the combination
\be
\hat\Pi_{-}=\tilde\eta\left[ \hat\S+i{\rm Im}(C\hat\O)\right]\label{Pidef}~.
\ee
The compensator $C$ \cite{Grimm:2004ua,Benmachiche:2006df} is given by
\be C=2e^{i\theta}e^{\frac{K_-}2-\varphi}~,\label{Ccomp}\ee
where the phase $e^{i\theta}$ identifies the orientifold projection, e.g. in the SU(3) structure case one has   
\be \hat\s^*J=-J~,\quad\quad\hat\s^*\O=e^{2i\theta}\bar{\O}\, . \ee
The operator $\hat\s^*$ is in fact  the pullback of the involutive symmetry $\s_o$ in the orientifold projector $\mathcal{O}\equiv \O_p(-1)^{F_L}\s_o$  \cite{Grimm:2004ua,Benmachiche:2006df}. The normalisation factor $\tilde\eta$ is defined as
\be\tilde\eta=e^{\frac{K_C-K_+}2}~,\label{etatilde}\ee
where
\be
e^{-K_C}=i \int \langle \bar{C}\hat{\bar\O},\,C\hat{\O}\rangle~.\\
\ee
Together these imply
\be 
e^{-K_C}=4e^{-2\varphi},\quad\quad\tilde\eta={\textstyle\frac12}e^{\varphi-\frac{K_+}2}~.
\ee
Note that since $\tilde\eta=\eta|C|^{-1}$ we can rewrite \eq{Pidef} as
\be
\hat\Pi_{-}=\left[\tilde\eta \hat\S+i{\rm Im}(e^{i\theta}\hat\Phi_-)\right]\label{p4}~.
\ee
In order to write ${\rm Re}\hat\Pi_-$ in a convenient way we introduce the rescaled RR scalars
\be 
(\z^A_{\tilde\eta},\,\tilde\z_{A\tilde\eta} )\equiv \tilde\eta (\z^A,\,\tilde\z_A)\label{Srescal}~.
\ee
Finally, we can define the rescaled RR flux as 
\be 
\hat F^{flux}_{\tilde\eta}\equiv\tilde\eta \hat F^{flux}=\tilde\eta\left(e_{\bf\L}\hat\o^{\bf\L}-m^{\bf\L}\hat\o_{\bf\L}\right)\,.\label{p5}
\ee


\subsubsection*{Non-geometric $Q$ and $R$ deformations}

In the $SU(3) \times SU(3)$ structure case with vanishing triplets the basis forms of the truncated spaces of forms are not necessarily of pure degree. This leads one to define a new generalised differential (see e.g. \cite{Grana:2006hr} and references therein)  
\be 
\mathcal{D}\equiv d\ -H\wedge\ \,-Q\cdot\ \,-R\llcorner\label{adefD}~,
\ee
where $Q\cdot\ $ and $R\llcorner\ $ act on a generic $k$-form $C$ as
\be 
(Q\cdot C)_{m_1\dots m_{k-1}}=Q^{ab}_{\ \ [m_1}C_{ab m_2\dots m_{k-1}]},\quad\quad (R\llcorner C)_{m_1\dots m_{k-3}}=R^{abc}C_{abcm_1\dots m_{k-3}}~.
\ee
The action of $\mathcal{D}$ on the basis forms gives 
\ba 
\mathcal{D}\hat\a_A=-m_A^{\bf\L}\hat\o_{\bf\L}+e_{A{\bf\L}}\hat\o^{\bf\L},\quad\quad \mathcal{D}\hat\b^A=-m^{A\bf\L}\hat\o_{\bf\L}+e^A_{\bf\L} \hat\o^{\bf\L} ~,\nn\\
\mathcal{D}\hat\o_{\bf\L}=e_{\bf\L}^A\hat\a_A-e_{A{\bf\L}}\hat\b^{A},\quad\quad \mathcal{D}\hat\o^{\bf\L}=m^{A\bf\L}\hat\a_A-m^{\bf\L}_A\hat\b^A\label{decohSU33}~,
\ea
while $\mathcal{D}^2=0$ implies
\ba  
&e_{A{\bf\L}}e^A_{~\bf\S}-e_{A{\bf\S}}e^A_{~\bf\L}=0,\quad\quad m_{A}^{\bf\L}m^{A\bf\S}-m_{A}^{\bf\S}m^{A\bf\L}=0&\nn ~,\\
\label{1notadSU33} &e_{A{\bf\L}}m^{A\bf\S}-e^A_{{\bf\S}}m_A^{\bf\L}=0~,&\\
&& \nn\\ 
&e_{A\bf\L}m^{B\bf\L}-e^B_{\bf\L}m_A^{\bf\L}=0,\quad\quad e_{A\bf\L}m_B^{\bf\L}-e_{B\bf\L}m_A^{\bf\L}=0&\nn~,\\
&e^A_{\bf\L}m^{B\bf\L}-e^B_{\bf\L}m^{A\bf\L}=0,\quad\quad e^A_{\bf\L}m_B^{\bf\L}-e_{B\bf\L}m^{A\bf\L}=0 &~.\label{2notadSU33} 
\ea
The parameters $(m_A^i,\,m^{Ai})$ are due to the action of the operator $Q\cdot\ $, while $(m_A^0,\,m^{0A})$ are due to the action of $R_\llcorner\ $.  We will refer to the former as $Q$ deformations and to the latter as $R$ deformations. While $Q$ deformations do not spoil the local geometric structure, the $R$ deformations do i.e. they are conjectured to arise from non-geometric compactifications.

Using the previous basis one finds
\ba
\label{dPhi+}\mathcal{D}\hat\Phi_+\!\!&=&\!\!\left[\left(X^{\bf\L}e_{\bf\L}^A-F_{\bf\L}\,m^{A\bf\L}\right)\hat\a_A-\left(X^{\bf\L}e_{A{\bf\L}}-F_{\bf\L}m^{\bf\L}_A\right)\hat\b^A\right]~,\\
\label{dPhi-}\mathcal{D}\hat\Phi_-\!\!&=&\!\!-\left[\left(Z^A_\eta m_A^{\bf\L}-G_{\eta A}\,m^{A\bf\L}\right)\hat\o_{\bf\L} - \left(Z^A_\eta e_{A{\bf\L}}-G_{\eta A}e^A_{\bf\L} \right)\hat\o^{\bf\L}\right]~,\\
\label{Ftorsflux}\hat{F}\!\!&=&\!\!-\left[\left(\z^A m_A^{\bf\L}-\tilde\z_{ A}\,m^{A\bf\L}\right)\hat\o_{\bf\L} - \left(\z^A e_{A{\bf\L}}-\tilde\z_{ A}e^A_{\bf\L} \right)\hat\o^{\bf\L}\right]~,\nn\\
\!\!&&\!\!-\left[m^{\bf\L}\hat\o_{\bf\L}-e_{\bf\L}\hat\o^{\bf\L}\right]~,
\ea
where we have defined the total RR contribution as
\be 
\hat F\equiv \mathcal{D}\hat\S+\hat F^{flux},\quad\quad \hat F_{\tilde\eta}\equiv\tilde\eta\hat F\label{RRtotal}~.
\ee
The $SU(3)$ structure case is recovered from \eq{dPhi+}-\eq{Ftorsflux} by setting $\mathcal{D}\rightarrow d_H$ and $m_A^{\bf\L}=m^{A\bf\L}=m^{\bf\L}=0$.


\section{From domain walls to Hitchin flows}\label{DWmanipul}

Here we shall reproduce this main steps of the calculations of section \ref{s2} that allow us to retrieve the Hitchin flow equations from the first-order differential equations describing domain wall configurations in four dimensions. The manipulation of \eq{attz0} is quite straightforward. We introduce for convenience the real symplectic vector $(W^{\bf\L},\,W_{\bf\L})$ 
\be
\label{Wlambda}W_{\bf\L}\equiv\vec n\cdot\vec P_{\bf\L}\,,\quad\quad W^{\bf\L}\equiv\vec n\cdot\vec Q^{\bf\L}~,
\ee
such that the superpotential \eq{SU2W} can be rewritten as
\be W= e^{\frac{K_+}2} \left(X^{\bf\L} W_{\bf\L}- F_{\bf\L} W^{\bf\L}\right)\ ,\label{Wredef}\ee
where explicitly 
\ba
W^{\bf\L}\!\!\!\!&=& \!\!\!\! 2 n^1\, e^{\varphi+\frac{K_-}2} \left[\left({\rm Re}G_A+\k |C|^{-1}\, \tilde\z_{A}\right) m^{A\L}-\left({\rm Re}Z^A+\textstyle{\frac{n^3}{n^1}|C|^{-1}}\,\z^A\right) m_{A}^{~\L}-\k |C|^{-1}\, m^\L\right]~,\nn\\
W_{\bf\L}\!\!\!\!&=&\!\!\!\!2n^1\, e^{\varphi+\frac{K_-}2} \left[\left({\rm Re}G_A+\k |C|^{-1}\,\tilde\z_{A}\right) e^A_{~\L}- \left({\rm Re}Z^A +\k |C|^{-1}\,\z^A e_{A\L}\right)-\k |C|^{-1}\, e_{\L}\right]~,\nn\\
\ea
and  $\k\equiv\frac{n^3}{n^1}$ is a constant according by \eq{Dn}. Assuming that $\vec{n}$ only depends on the quaternionic scalars, the vectors \eq{Wlambda} are independent of the $t^i$ and we can rewrite equation \eq{attz0} as
\be\label{attznew} 
\partial_rt^i=\mp e^{-pU}g^{i\bar{\jmath}}\left[\partial_{\bar\jmath}(e^{\frac{K_+}2}\hb \bar X^{\bf\L})W_{\bf\L}-\partial_{\bar\jmath}(e^{\frac{K_+}2}\hb \bar F_{\bf\L})W^{\bf\L}\right]~.
\ee
We can now proceed exactly as in \cite{Louis:2006wq}, as the specific structure of \eq{Wlambda} plays no role in the manipulations.

Following the procedure outlined in \cite{Behrndt:2001mx}, we introduce the symplectic section
\be 
\label{newsect}\begin{pmatrix} Y^{\bf\L}\cr
\mathcal{F}_{\bf\L} \end{pmatrix} \equiv h\,e^{U(r)}e^{\frac{K_+}{2}}\begin{pmatrix} X^{\bf\L}\cr
F_{\bf\L} \end{pmatrix} ~,
\ee
and from \eq{attznew} we obtain
\be
\partial_r\begin{pmatrix} Y^{\bf\L}-\overline{Y}^{\bf\L}\cr \mathcal{F}_{\bf\L}-\overline{\mathcal{F}}_{\bf\L}  \end{pmatrix} =-ie^{(1-p)U}\begin{pmatrix} W^{\bf\L}\cr W_{\bf\L} \end{pmatrix} \label{dRephi+}~.
\ee
Note that \eq{dRephi+} can be formally integrated to give 
\be
\begin{pmatrix} {\rm Im}(hX^{\bf\L})\cr {\rm Im }
(hF_{\bf\L}) \end{pmatrix} =\begin{pmatrix} \mathcal{I}_Am^{A\bf\L}-\mathcal{I}^Am_A^{\bf\L}-\mathcal{I}m^{\bf\L}\cr \mathcal{I}_Ae^A_{\bf\L}-\mathcal{I}^Ae_{A\bf\L}-\mathcal{I}e_{\bf\L} \end{pmatrix} \label{intat}~,
\ee
where $(\mathcal{I}^A,\,\mathcal{I}_A)$ and $\mathcal{I}$ result from integration and their explicit form is not relevant for our purposes. Plugging \eq{intat} into \eq{dPhi+} and using \eq{2notadSU33} we deduce
\be 
d{\rm Im}(h\Phi^0_+)=0~.
\ee
Another important consequence of  \eq{intat} and \eq{2notadSU33} is that
\be 
{\rm Im}(h\vec W)=0\label{ImhvecW}~.
\ee
Note that if we multiply \eq{ImhvecW} by $\vec n$ we obtain \eq{reW}. Moreover, \eq{intat} and \eq{2notadSU33} imply that only $\left({\rm Re}(hX^\Lambda),\,{\rm Re}(hF_\Lambda)\right)$ appear 
in $h\vec{W}$.

Equation\eq{attq0} is more difficult to manipulate. Actually, we are just interested in the scalars  $(z^a,\bar{z}^{\bar a})$ in the special K\"ahler submanifold parameterising  the complex structure deformations.  Rewriting \eq{attq0} in complex coordinates and  assuming that $W^2=0$, we obtain
\be
\partial_r z^a=\mp e^{-pU} g^{a\bar b}\left[\nabla_{\bar b}(e^{\frac{K_-}2}\bar Z^A)\mathcal{E}_A- \nabla_{\bar b}(e^{\frac{K_-}2}\bar G_A)\mathcal{E}^A\right]\label{attSKQ}~,
\ee 
with
\be
\begin{pmatrix} \mathcal{E}^A\cr\mathcal{E}_A \end{pmatrix} =2n^1\,e^{\varphi+\frac{K_+}2} \begin{pmatrix}  {\rm Re}(hF)_{\bf\Lambda}\, m^{A{\bf\L}}- {\rm Re}(hX^{\bf \Lambda})\, e^A_{\bf\L} \cr  {\rm Re}(hF)_{\bf \Lambda}\, m_{A}^{\bf\L}- {\rm Re}(hX^{\bf \Lambda})\, e_{A\bf\L} \end{pmatrix} \label{WC}~.
\ee
As \eq{attSKQ} has the same structure as \eq{attznew} it can be manipulated with similar techniques. Let us make some comments on the derivation of \eq{attSKQ}. First of all, observe that due to the structure of the quaternionic metric \cite{FeSab} the equations for ${\rm Re}z^a$ and ${\rm Im}z^a$ decouple from the rest. Secondly, note that $D_aW^3=0$, since $W^3$ does note depend on  $({\rm Re}z^a,\,{\rm Im}z^a)$ (see \eq{WSU3} and \eq{WSU33}) and that the connection components $\o^1,\,\o^2$ are zero along the directions $({\rm Re}z^a,\,{\rm Im}z^a)$  \cite{FeSab}. Thirdly, we used that $e^{\frac{K_-}2}(Z^A,\,G_A)$ is a covariantly holomorphic section. Finally, we imposed $W^2=0$ (and consequently $n^2=0$) in order to remove a contribution which would have spoiled the structure of \eq{attSKQ}. Introducing, as before, a new symplectic section

\be 
\label{newsect2}\begin{pmatrix} \mathcal Y^A\cr
\mathcal{G}_A \end{pmatrix} \equiv e^{(1-\lambda)U(r)}e^{\frac{K_-}{2}}\begin{pmatrix} Z^A\cr
G_A \end{pmatrix} ~,
\ee
we obtain
\be
\partial_r\begin{pmatrix} \mathcal Y^A-\overline{\mathcal Y}^A\cr \mathcal{G}_A-\overline{\mathcal{G}}_A  \end{pmatrix} =-ie^{(1-\lambda-p)U}\begin{pmatrix} \mathcal{E}^A\cr\mathcal{E}_A \end{pmatrix} \label{attquat}~,
\ee
where we have defined $\lambda= (n^3)^2$ and made use of \eq{normphase}. We can formally integrate \eq{attquat} to obtain
\be
\begin{pmatrix} {\rm Im}Z^A\cr{\rm Im} G_A \end{pmatrix} =\begin{pmatrix} \mathcal{I}^{\bf\L}\,e^A_{\bf\L}-\mathcal{I}_{\bf\L}\,m^{A{\bf\L}}\cr \mathcal{I}^{\bf\L}\,e_{A\bf\L}-\mathcal{I}_{\bf\L}\,m_{A}^{\bf\L} \end{pmatrix} \label{intatt}~,
\ee
where $(\mathcal{I}^{\bf\L},\,\mathcal{I}_{\bf\L})$ again result from the integration.  Plugging \eq{intatt} into \eq{dPhi-}, we immediately see that, by virtue of \eq{1notadSU33}, we have
\be
d{\rm Im}\Phi_{-}^0=0~,\label{W2=0}
\ee 
which is consistent with $W^2=0$.  We stress that the assumption $W^2=0$ is necessary to get an attractor-like structure \eq{attquat} for the complex structure deformations analogous to \eq{dRephi+} for the K\"ahler cone.

Considering  \eq{attq0} for $q^u=\varphi$, together with \eq{attu}, we can obtain the following expression for the four-dimensional dilaton
\be
e^{\varphi}= e^{-(1+\lambda)U}\label{4dimdil}~,
\ee
where we have used \cite{FeSab} $g^{\varphi\varphi}=1$ and \eq{normphase} with the definitions $\k\equiv\frac{n3}{n1}$ and $\l\equiv\frac{\k2}{1+\k2}$.


\begin{thebibliography}{99}

\bibitem{Grana:2005jc}
M.~Grana, ``Flux compactifications in string theory: A comprehensive review'', Phys.\ Rept.\  {\bf 423} (2006) 91 [{\tt hep-th/0509003}].

\bibitem{Hitchin:2004ut}
N.~Hitchin, ``Generalized Calabi-Yau manifolds'', Quart.\ J.\ Math.\ Oxford Ser.\  {\bf 54} (2003) 281 [{\tt math/0209099}].

\bibitem{Gualtieri:2003dx}
 M.~Gualtieri,  ``Generalized complex geometry'', [{\tt math/0401221}].

\bibitem{gmpt1}
M.~Grana, R.~Minasian, M.~Petrini and A.~Tomasiello,
``Supersymmetric backgrounds from generalised Calabi-Yau manifolds'',
JHEP {\bf 0408} (2004) 046 [{\tt hep-th/0406137}].

\bibitem{Witt}
C.~Jeschek and F.~Witt, ``Generalised G(2)-structures and type IIB superstrings'', JHEP {\bf 0503} (2005) 053 [{\tt hep-th/0412280}].

\bibitem{gmpt2}
M.~Grana, R.~Minasian, M.~Petrini and A.~Tomasiello,  ``Generalised structures of N = 1 vacua'', JHEP {\bf 0511} (2005) 020 [{\tt hep-th/0505212}].

\bibitem{gmpt3}
M.~Grana, R.~Minasian, M.~Petrini and A.~Tomasiello, ``A scan for new N=1 vacua on twisted tori'',
 JHEP {\bf 0705} (2007) 031 [{\tt hep-th/0609124}].

\bibitem{Grana:2005ny}
M.~Grana, J.~Louis and D.~Waldram, ``Hitchin functionals in N = 2 supergravity'', JHEP {\bf 0601} (2006) 008 [{\tt hep-th/0505264}].

\bibitem{Grana:2006hr}
M.~Grana, J.~Louis and D.~Waldram, ``SU(3) x SU(3) compactification and mirror duals of magnetic fluxes'', JHEP {\bf 0704} (2007) 101
[{\tt hep-th/0612237}].

\bibitem{Tomasiello:2007eq}
A.~Tomasiello, ``New string vacua from twistor spaces'', Phys.\ Rev.\  D {\bf 78} (2008) 046007 [{\tt 0712.1396 [hep-th]}].

\bibitem{Kounnas:2007dd}
C.~Kounnas, D.~Lust, P.~M.~Petropoulos and D.~Tsimpis, ``AdS4 flux vacua in type II superstrings and their domain-wall solutions'', JHEP {\bf 0709} (2007) 051 [{\tt 0707.4270 [hep-th]}].  

\bibitem{Koerber:2008rx}
P.~Koerber, D.~Lust and D.~Tsimpis, ``Type IIA AdS4 compactifications on cosets, interpolations and domain walls'', JHEP {\bf 0807} (2008) 017 [{\tt 0804.0614 [hep-th]}].

\bibitem{Kachru:2003aw}
S.~Kachru, R.~Kallosh, A.~Linde and S.~P.~Trivedi, ``De Sitter vacua in string theory'', Phys.\ Rev.\  D {\bf 68} (2003) 046005 [{\tt hep-th/0301240}].

\bibitem{Mayer:2004sd}
C.~Mayer and T.~Mohaupt, ``Domain walls, Hitchin's flow equations and G(2)-manifolds'', Class.\ Quant.\ Grav.\  {\bf 22} (2005) 379 [{\tt hep-th/0407198}]. 

\bibitem{Louis:2006wq} J.~Louis and S.~Vaula, ``N = 1 domain wall solutions of massive type II supergravity as generalised
 geometries'', JHEP {\bf 0608} (2006) 058 [{\tt hep-th/0605063}].  
 
\bibitem{Behrndt:2001mx}
K.~Behrndt, G.~Lopes Cardoso and D.~Lust, ``Curved BPS domain wall solutions in four-dimensional N = 2  supergravity'', Nucl.\ Phys.\  B {\bf 607} (2001) 391 [{\tt hep-th/0102128}]. 

\bibitem{Lust:2008zd}
D.~Lust, F.~Marchesano, L.~Martucci and D.~Tsimpis,``Generalized non-supersymmetric flux vacua'', JHEP {\bf 0811} (2008) 021 [{\tt 0807.4540 [hep-th]}].

\bibitem{Cvetic:1996vr}
M.~Cvetic and H.~H.~Soleng, ``Supergravity domain walls'', Phys.\ Rept.\  {\bf 282} (1997) 159 [{\tt hep-th/9604090}].

\bibitem{HitchinHF}
N. Hitchin, ``Stable forms and special metrics'',
in ``Global Differential Geometry: The Mathematical Legacy of Alfred
Gray'', M.Fernandez and J.A.Wolf (eds.), 
Contemporary Mathematics {\bf 288}, American Mathematical Society,
Providence (2001) [{\tt math.DG/0107101}]. 

\bibitem{Jeschek}
C.~Jeschek and F.~Witt, ``Generalised geometries, constrained critical points and Ramond-Ramond fields'', [{\tt math/0510131}].

\bibitem{Witt2}  
F.~Witt, ``Generalised $G_2$-manifolds'', Commun.\ Math.\ Phys.\  {\bf 265} (2006) 275 [{\tt  math/0411642}].

\bibitem{KashaniPoor:2007tr}
 A.~K.~Kashani-Poor, ``Nearly Kaehler Reduction'', JHEP {\bf 0711} (2007) 026 [{\tt 0709.4482 [hep-th]}].

\bibitem{Caviezel:2008ik}
C.~Caviezel, P.~Koerber, S.~Kors, D.~Lust, D.~Tsimpis and M.~Zagermann, ``The effective theory of type IIA AdS4 compactifications on nilmanifolds and cosets'', JHEP {\bf 0807} (2008) 017 [{\tt 0806.3458 [hep-th]}].  
 
\bibitem{Dall'Agata:2003yr}
G.~Dall'Agata, R.~D'Auria, L.~Sommovigo and S.~Vaula, ``D = 4, N = 2 gauged supergravity in the presence of tensor multiplets'',  Nucl.\ Phys.\  B {\bf 682} (2004) 243, [{\tt hep-th/0312210}]. R.~D'Auria, L.~Sommovigo and S.~Vaula, ``N = 2 supergravity Lagrangian coupled to tensor multiplets with  electric and magnetic fluxes'', JHEP {\bf 0411} (2004) 028 [{\tt hep-th/0409097}].

\bibitem{Sommovigo:2004vj}
L.~Sommovigo and S.~Vaula, ``D = 4, N = 2 supergravity with Abelian electric and magnetic charge'', Phys.\ Lett.\  B {\bf 602} (2004) 130  [{\tt hep-th/0407205}].
  
\bibitem{D'Auria:2004tr}
R.~D'Auria, S.~Ferrara, M.~Trigiante and S.~Vaula, ``Gauging the Heisenberg algebra of special quaternionic manifolds'', Phys.\ Lett.\  B {\bf 610} (2005) 147, [{\tt hep-th/0410290}]. R.~D'Auria, S.~Ferrara, M.~Trigiante and S.~Vaula, ``Scalar potential for the gauged Heisenberg algebra and a non-polynomial antisymmetric tensor theory'', Phys.\ Lett.\  B {\bf 610} (2005) 270 [{\tt hep-th/0412063}].

\bibitem{D'Auria:2007ay}
R.~D'Auria, S.~Ferrara and M.~Trigiante, ``On the supergravity formulation of mirror symmetry in generalised Calabi-Yau manifolds'', Nucl.\ Phys.\  B {\bf 780} (2007) 28 [{\tt hep-th/0701247}].

\bibitem{LopesCardoso:2001rt}
G.~Lopes Cardoso, G.~Dall'Agata and D.~Lust,
 ``Curved BPS domain wall solutions in five-dimensional  gauged supergravity'',
 JHEP {\bf 0107} (2001) 026
 [{\tt hep-th/0104156}]. 
 
\bibitem{hmt}
J.~P.~Hsu, A.~Maloney and A.~Tomasiello, ``Black hole attractors and pure spinors'', JHEP {\bf 0609} (2006) 048 [{\tt hep-th/0602142}].

\bibitem{hlmt}
M.~Haack, D.~{L\"ust}, L.~Martucci, and A.~Tomasiello, [{\tt 0905.1582[hep-th]}].

\bibitem{Cassani:2009ck}
D.~Cassani and A.~K.~Kashani-Poor, ``Exploiting N=2 in consistent coset reductions of type IIA'', [{\tt 0901.4251 [hep-th]}].

\bibitem{Andrianopoli:2001gm}
 L.~Andrianopoli, R.~D'Auria and S.~Ferrara,
 ``Consistent reduction of N = 2 $\rightarrow$ N = 1 four dimensional supergravity coupled to matter'',
 Nucl.\ Phys.\  B {\bf 628} (2002) 387
 [{\tt hep-th/0112192}].

\bibitem{D'Auria:2005yg}
 R.~D'Auria, S.~Ferrara, M.~Trigiante and S.~Vaula,
 ``N = 1 reductions of N = 2 supergravity in the presence of tensor multiplets'',
 JHEP {\bf 0503} (2005) 052
 [{\tt hep-th/0502219}].

\bibitem{Cassani:2007pq}
D.~Cassani and A.~Bilal, ``Effective actions and N=1 vacuum conditions from SU(3) x SU(3) compactifications'', JHEP {\bf 0709} (2007) 076 [{\tt 0707.3125 [hep-th]}].

\bibitem{Grimm:2004ua}
T.~W.~Grimm and J.~Louis, ``The effective action of type IIA Calabi-Yau orientifolds'', Nucl.\ Phys.\  B {\bf 718} (2005) 153 [{\tt hep-th/0412277}].

 
\bibitem{Benmachiche:2006df}
I.~Benmachiche and T.~W.~Grimm, ``Generalized N = 1 orientifold compactifications and the Hitchin functionals'',  Nucl.\ Phys.\  B {\bf 748} (2006) 200   [{\tt hep-th/0602241}].

\bibitem{Andriot:2009fp}
D.~Andriot, R.~Minasian and M.~Petrini, ``Flux backgrounds from Twist duality'', [{\tt 0903.0633 [hep-th]}]. 

\bibitem{bkorp}
 E.~Bergshoeff, R.~Kallosh, T.~Ortin, D.~Roest and A.~Van Proeyen,
 ``New formulations of D = 10 supersymmetry and D8 - O8 domain walls'',
 Class.\ Quant.\ Grav.\  {\bf 18} (2001) 3359
 [{\tt hep-th/0103233}]. 
 

\bibitem{ms}
L.~Martucci and P.~Smyth, ``Supersymmetric D-branes and calibrations on general N = 1 backgrounds'',  JHEP {\bf 0511} (2005) 048 [{\tt hep-th/0507099}]. 

\bibitem{Koerber:2005qi}
P.~Koerber, ``Stable D-branes, calibrations and generalized Calabi-Yau geometry'', JHEP {\bf 0508} (2005) 099 [{\tt hep-th/0506154}].
 

\bibitem{Aharony:2008wz}
O.~Aharony, Y.~E.~Antebi and M.~Berkooz,``On the Conformal Field Theory Duals of type IIA $AdS_4$ Flux Compactifications'', JHEP {\bf 0802} (2008) 093
[{\tt 0801.3326 [hep-th]}]. 

\bibitem{ten2four}
P.~Koerber and L.~Martucci, ``From ten to four and back again: how to generalise the geometry'', JHEP {\bf 0708} (2007) 059 [{\tt 0707.1038 [hep-th]}].

\bibitem{adsbranes}
P.~Koerber and L.~Martucci, ``D-branes on AdS flux compactifications'', JHEP {\bf 0801} (2008) 047 [{\tt 0710.5530 [hep-th]}].

\bibitem{Koerber:2007hd} P.~Koerber and D.~Tsimpis, ``Supersymmetric sources, integrability and generalised-structure compactifications'', JHEP {\bf 0708} (2007) 082 [{\tt 0706.1244 [hep-th]}]. 
  
\bibitem{SV2}
P.~Smyth and S.~Vaula, ``Warped compactifications and domain walls'', work in progress. 
     
\bibitem{Gurrieri:2002wz}
S.~Gurrieri, J.~Louis, A.~Micu and D.~Waldram, ``Mirror symmetry in generalised Calabi-Yau compactifications'', Nucl.\ Phys.\  B {\bf 654} (2003) 61   [{\tt hep-th/0211102}].  

\bibitem{pk} P.~Koerber, private communication.

\bibitem{Gaiotto:2009yz}
D.~Gaiotto and A.~Tomasiello, ``Perturbing gauge/gravity duals by a Romans mass'', [{\tt 0904.3959 [hep-th]}].

\bibitem{Petrini:2009ur}
M.~Petrini and A.~Zaffaroni, ``N=2 solutions of massive type IIA and their Chern-Simons duals'', [{\tt 0904.4915 [hep-th]}].

\bibitem{Andrianopoli:1996cm}
For a review see, for example, 
L.~Andrianopoli, M.~Bertolini, A.~Ceresole, R.~D'Auria, S.~Ferrara, P.~Fre and T.~Magri,
``N = 2 supergravity and N = 2 super Yang-Mills theory on general scalar
manifolds: Symplectic covariance, gaugings and the momentum map'',
J.\ Geom.\ Phys.\  {\bf 23} (1997) 111
[{\tt hep-th/9605032}] and references therein.

\bibitem{CS} S.~Chiossi and S.~Salamon, ``The Intrinsic Torsion of SU(3) and G2 Structures'', 
Differential geometry, Valencia, 2001, pp. 115, [{\tt math.DG/0202282}]. 

\bibitem{FeSab}
S.~Ferrara and S.~Sabharwal, ``Quaternionic Manifolds for Type II Superstring Vacua of Calabi-Yau Spaces'', Nucl.\ Phys.\  B {\bf 332} (1990) 317.


\end{thebibliography}
\end{document}